\let\TeXyear\year
\documentclass[pageno]{jpaper}

\let\year\TeXyear
\usepackage{cite}
\usepackage{amsmath, amsfonts}
\usepackage[noend]{algorithmic}
\usepackage{graphicx}
\usepackage{adjustbox}
\usepackage{textcomp}

\usepackage{xspace}
\newcommand{\ptxt}{{plaintext}\xspace}
\newcommand{\ctxt}{{ciphertext}\xspace}
\newcommand{\ptxts}{{plaintexts}\xspace}
\newcommand{\ctxts}{{ciphertexts}\xspace}

\usepackage{indentfirst}
\usepackage{soul}
\usepackage{tabularx}
\usepackage{microtype}
\usepackage{booktabs, tabularx}
\usepackage{url}
\usepackage{xcolor, soul}
\newcommand*{\Scale}[2][4]{\scalebox{#1}{$#2$}}
\usepackage{multirow} 
\usepackage{dcolumn}
\newcolumntype{d}[1]{D{.}{.}{#1}}
\newcommand\mc[1]{\multicolumn{1}{c}{#1}} 
\newcommand{\reviewertwo}[1]{{\color{black}#1}}
\newcommand{\reviewerthree }[1]{{\color{black}#1}}
\soulregister\cite7
\soulregister\ref7
\soulregister\ptxts7
\soulregister\ctxts7
\soulregister\ptxt7
\soulregister\ctxt7

\usepackage{mathtools}
\usepackage{enumitem}
\setlist[itemize]{leftmargin=*, itemsep=1.5pt, topsep=3pt}
\setlist[enumerate]{leftmargin=*, itemsep=1.5pt, topsep=3pt}
\usepackage{wrapfig}
\usepackage{comment}
\usepackage[normalem]{ulem}

\usepackage{tikz} 
\usetikzlibrary{patterns,tikzmark, arrows}
\usetikzlibrary{matrix,decorations.pathreplacing,calc,fit,backgrounds,calligraphy} 
\usetikzlibrary{external}
\newcommand{\myhash}{|[fill=gray!20]|{\scalebox{0.7}{\#\#}}}
\newcommand{\myzero}{|[fill=gray!20]|0}

\tikzset{
    toprule/.style={%
        execute at end cell={%
            \draw [line cap=rect,#1] (\tikzmatrixname-\the\pgfmatrixcurrentrow-\the\pgfmatrixcurrentcolumn.north west) -- (\tikzmatrixname-\the\pgfmatrixcurrentrow-\the\pgfmatrixcurrentcolumn.north east);%
        }
    },
    bottomrule/.style={%
        execute at end cell={%
            \draw [line cap=rect,#1] (\tikzmatrixname-\the\pgfmatrixcurrentrow-\the\pgfmatrixcurrentcolumn.south west) -- (\tikzmatrixname-\the\pgfmatrixcurrentrow-\the\pgfmatrixcurrentcolumn.south east);%
        }
    }
}

\usetikzlibrary{matrix, positioning}
\pgfmathsetmacro{\myscale}{1}
\pgfkeys{tikz/mymatrixenv/.style={decoration={brace},every left delimiter/.style={xshift=8pt},every right delimiter/.style={xshift=-8pt}}}
\pgfkeys{tikz/mymatrix/.style={matrix of math nodes,nodes in empty cells,
left delimiter={[},right delimiter={]},inner sep=1pt,outer sep=1.5pt,
column sep=8pt,row sep=8pt,nodes={minimum width=20pt,minimum height=10pt,
anchor=center,inner sep=0pt,outer sep=0pt,scale=\myscale,transform shape}}}
\pgfkeys{tikz/mymatrixbrace/.style={decorate,thick}}

\tikzset{greenish/.style={
    fill=green!50!lime!60,draw opacity=0.4,
    draw=green!50!lime!60,fill opacity=0.1,
  },
  cyanish/.style={
    fill=cyan!90!blue!60, draw opacity=0.4,
    draw=blue!70!cyan!30,fill opacity=0.1,
  },
  orangeish/.style={
    fill=orange!90, draw opacity=0.8,
    draw=orange!90, fill opacity=0.3,
  },
  brownish/.style={
    fill=brown!70!orange!40, draw opacity=0.4,
    draw=brown, fill opacity=0.3,
  },
  purpleish/.style={
    fill=violet!90!pink!20, draw opacity=0.5,
    draw=violet, fill opacity=0.3,    
  }}

\tikzset{c/.style={fill=yellow!10}}
\tikzset{mystyle/.style={matrix of nodes,
        nodes in empty cells,
        nodes={draw, fill=teal!30},
        row sep=-\pgflinewidth,
        column sep=-\pgflinewidth,
        nodes={draw,minimum width=0.75cm,minimum height=0.75cm,anchor=center}}}

\tikzset{mystyle1by3/.style={matrix of nodes,
        nodes in empty cells,
        nodes={draw, fill=teal!30},
        row sep=-\pgflinewidth,
        column sep=-\pgflinewidth,
        font = \small,
        column 3/.style={column sep=5pt},
        column 2/.style={nodes ={minimum width=1.5cm}},
        column 5/.style={nodes ={minimum width=1.5cm}},
        nodes={draw,minimum width=0.75cm,minimum height=0.75cm,anchor=center}}}

\tikzset{mystyle1by2/.style={matrix of nodes,
        nodes in empty cells,
        nodes={draw, fill=teal!30},
        row sep=-\pgflinewidth,
        column sep=-\pgflinewidth,
        font = \small,
        column 2/.style={nodes ={minimum width=1.5cm},column sep=5pt},
        column 4/.style={nodes ={minimum width=1.5cm}},
        nodes={draw,minimum width=2.2cm,minimum height=0.75cm,anchor=center}}}

\tikzset{mystyle1/.style={matrix of nodes,
        nodes in empty cells,
        nodes={draw, fill=blue!20},
        row sep=-\pgflinewidth,
        column sep=-\pgflinewidth,
        font = \small,
        nodes={draw,minimum width=0.75cm,minimum height=0.75cm,anchor=center}}}

\tikzset{mystyle11/.style={matrix of nodes,
        nodes in empty cells,
        nodes={draw, fill=mygreen!25},
        row sep=-\pgflinewidth,
        column sep=-\pgflinewidth,
        nodes={draw,minimum width=0.88cm,minimum height=0.75cm,anchor=center}
        }}

\tikzset{mystyle111/.style={matrix of nodes,
        nodes in empty cells,
        nodes={draw, fill=brown!30},
        row sep=-\pgflinewidth,
        column sep=-\pgflinewidth,
        nodes={draw,minimum width=0.88cm,minimum height=0.75cm,anchor=center}}}

\tikzset{mystyle111/.style={matrix of nodes,
        nodes in empty cells,
        nodes={draw, fill=brown!30},
        row sep=-\pgflinewidth,
        column sep=-\pgflinewidth,
        nodes={draw,minimum width=0.88cm,minimum height=0.75cm,anchor=center}}}

\tikzset{mystyle2/.style={matrix of nodes,
        nodes in empty cells,
        nodes={draw, fill=teal!30},
        row sep=-\pgflinewidth,
        column sep=-\pgflinewidth,
        font = \small,
        nodes={draw,minimum width=0.75cm,minimum height=0.75cm,anchor=center}}}

\tikzset{mystyle2_mid/.style={matrix of nodes,
        nodes in empty cells,
        nodes={draw, fill=blue!20},
        row sep=-\pgflinewidth,
        column sep=-\pgflinewidth,
        font = \small,
        nodes={draw,minimum width=1.0cm,minimum height=0.75cm,anchor=center}}}

\tikzset{mystyle2_long/.style={matrix of nodes,
        nodes in empty cells,
        nodes={draw, fill=blue!20},
        row sep=-\pgflinewidth,
        column sep=-\pgflinewidth,
        font = \small,
        column 1/.style={nodes ={minimum width=2.2cm}},
        column 3/.style={nodes ={minimum width=2.2cm}},
        column 2/.style={nodes ={minimum width=1.5cm},column sep=5pt},
        column 4/.style={nodes ={minimum width=1.5cm}},
        nodes={draw,minimum width=0.75cm,minimum height=0.75cm,anchor=center}}}

\tikzset{style3/.style={matrix of nodes,
        nodes in empty cells,
        nodes={draw, fill=orange!30},
        row sep=-\pgflinewidth,
        column sep=-\pgflinewidth,
        nodes={draw,minimum width=0.75cm,minimum height=0.75cm,anchor=center}}}

\tikzset{style33/.style={matrix of nodes,
        nodes in empty cells,
        nodes={draw, fill=red!20},
        row sep=-\pgflinewidth,
        column sep=-\pgflinewidth,
        nodes={draw,minimum width=0.88cm,minimum height=0.75cm,anchor=center}}}

\tikzset{style333/.style={matrix of nodes,
        nodes in empty cells,
        nodes={draw, fill=red!20},
        row sep=-\pgflinewidth,
        column sep=-\pgflinewidth,
        nodes={draw,minimum width=0.75cm,minimum height=0.75cm,anchor=center}}}

\tikzset{mystyle3by3/.style={matrix of nodes,
        nodes in empty cells,
        nodes={draw, fill=orange!30},
        row sep=-\pgflinewidth,
        column sep=-\pgflinewidth,
        column 1/.style={nodes ={minimum width=2.5cm}},
        column 2/.style={nodes ={minimum width=1cm},minimum height=0.75cm},
        row 1/.style={ nodes={draw, fill=orange!40}},
        row 2/.style={ nodes={draw, fill=orange!10}},
        row 3/.style={ nodes={draw, fill=orange!10}},
        row 4/.style={ nodes={draw, fill=orange!10}},
        column 3/.style={nodes ={minimum width=2.5cm},column sep=10pt},
        column 4/.style={nodes ={minimum width=2cm}},
        column 5/.style={nodes ={minimum width=1cm, minimum height=0.75cm}},
        column 6/.style={nodes ={minimum width=2.4cm}},
        nodes={draw,minimum width=2cm,minimum height=0.75cm,anchor=center}}}
\tikzset{mystyle2by2/.style={matrix of nodes,
        nodes in empty cells,
        nodes={draw, fill=blue!20},
        row sep=-\pgflinewidth,
        column sep=-\pgflinewidth,
        font = \small,
        column 2/.style={nodes ={minimum width=1cm},column sep=5pt},
        column 4/.style={nodes ={minimum width=1cm}},
        nodes={draw,minimum width=1.5cm,minimum height=0.75cm,anchor=center}}}
        
\tikzset{mystyle3by2/.style={matrix of nodes,
        nodes in empty cells,
        nodes={draw, fill=blue!20},
        row sep=-\pgflinewidth,
        column sep=-\pgflinewidth,
        font = \small,
        column 3/.style={nodes ={minimum width=1cm},column sep=5pt},
        column 6/.style={nodes ={minimum width=1cm}},
        nodes={draw,minimum width=1.5cm,minimum height=0.75cm,anchor=center}}}

\tikzset{mystyle1by2_long/.style={matrix of nodes,
        nodes in empty cells,
        nodes={draw, fill=blue!20},
        row sep=-\pgflinewidth,
        column sep=-\pgflinewidth,
        font = \small,
        column 2/.style={nodes ={minimum width=1.5cm},column sep=5pt},
        column 4/.style={nodes ={minimum width=1.5cm}},
        nodes={draw,minimum width=2.9cm,minimum height=0.75cm,anchor=center}}}

\tikzset{im2colctxt/.style={matrix of nodes,
        nodes in empty cells,
        nodes={draw, fill=teal!20},
        row sep=-\pgflinewidth,
        column sep=-\pgflinewidth,
        column 1/.style={nodes ={minimum width=2.5cm}},
        column 2/.style={nodes ={minimum width=1cm},minimum height=0.75cm},
        column 3/.style={nodes ={minimum width=2.5cm},column sep=10pt},
        column 4/.style={nodes ={minimum width=2cm}},
        column 5/.style={nodes ={minimum width=1cm, minimum height=0.75cm}},
        column 6/.style={nodes ={minimum width=2.4cm}},
        nodes={draw,minimum width=2cm,minimum height=0.01cm,anchor=center}}}
        
\tikzset{im2colweight/.style={matrix of nodes,
        nodes in empty cells,
        nodes={draw, fill=orange!30},
        row sep=-\pgflinewidth,
        column sep=-\pgflinewidth,
        column 1/.style={nodes ={minimum width=2.5cm}},
        column 2/.style={nodes ={minimum width=1cm},minimum height=0.75cm},
        column 3/.style={nodes ={minimum width=2.5cm},column sep=10pt},
        column 4/.style={nodes ={minimum width=2cm}},
        column 5/.style={nodes ={minimum width=1cm, minimum height=0.75cm}},
        column 6/.style={nodes ={minimum width=2.4cm}},
        nodes={draw,minimum width=2cm,minimum height=0.01cm,anchor=center}}}
        
\tikzset{im2coloutput/.style={matrix of nodes,
        nodes in empty cells,
        nodes={draw, fill=blue!20},
        row sep=-\pgflinewidth,
        column sep=-\pgflinewidth,
        column 1/.style={nodes ={minimum width=2.5cm}},
        column 2/.style={nodes ={minimum width=1cm},minimum height=0.75cm},
        column 3/.style={nodes ={minimum width=2.5cm},column sep=10pt},
        column 4/.style={nodes ={minimum width=2cm}},
        column 5/.style={nodes ={minimum width=1cm, minimum height=0.75cm}},
        column 6/.style={nodes ={minimum width=2.4cm}},
        nodes={draw,minimum width=2cm,minimum height=0.01cm,anchor=center}}}

\tikzset{mystyle2by3/.style={matrix of nodes,
        nodes in empty cells,
        nodes={draw, fill=teal!30},
        row sep=-\pgflinewidth,
        column sep=-\pgflinewidth,
        font = \small,
        column 3/.style={column sep=5pt},
        column 2/.style={nodes ={minimum width=1cm}},
        column 5/.style={nodes ={minimum width=1cm}},
        nodes={draw,minimum width=0.75cm,minimum height=0.75cm,anchor=center}}}

\tikzset{mystyle2by3_long/.style={matrix of nodes,
        nodes in empty cells,
        nodes={draw, fill=blue!20},
        row sep=-\pgflinewidth,
        column sep=-\pgflinewidth,
        font = \small,
        row 1/.style={nodes={draw, fill=blue!40}},
        row 2/.style={text opacity =1.0, opacity=0.5},
        row 3/.style={text opacity =1.0,opacity=0.5},
        row 4/.style={text opacity =1.0,opacity=0.5},
        column 3/.style={column sep=5pt},
        column 2/.style={nodes ={minimum width=1cm}},
        column 5/.style={nodes ={minimum width=1cm}},
        nodes={draw,minimum width=2.6cm,minimum height=0.75cm,anchor=center}}}

\tikzset{mystyle2by3_longlong/.style={matrix of nodes,
        nodes in empty cells,
        nodes={draw, fill=teal!30},
        row sep=-\pgflinewidth,
        column sep=-\pgflinewidth,
        font = \small,
        column 3/.style={column sep=5pt},
        column 2/.style={nodes ={minimum width=1.5cm}},
        column 5/.style={nodes ={minimum width=1.5cm}},
        nodes={draw,minimum width=6.2cm,minimum height=0.75cm,anchor=center}}}

\tikzset{style2/.style={matrix of nodes,
        nodes in empty cells,
        nodes={draw, fill=blue!20},
        row sep=-\pgflinewidth,
        column sep=-\pgflinewidth,
        nodes={draw,minimum width=0.75cm,minimum height=0.75cm,anchor=center}}}

\tikzset{style2_mid/.style={matrix of nodes,
        nodes in empty cells,
        nodes={draw, fill=blue!20},
        row sep=-\pgflinewidth,
        column sep=-\pgflinewidth,
        nodes={draw,minimum width=0.8cm,minimum height=0.75cm,anchor=center}}}

\tikzset{style22/.style={matrix of nodes,
        nodes in empty cells,
        nodes={draw, fill=cyan!40},
        row sep=-\pgflinewidth,
        column sep=-\pgflinewidth,
        nodes={draw,minimum width=0.75cm,minimum height=0.75cm,anchor=center}}}

\newcommand*\mymatrixbraceright[4][]{
    \draw[mymatrixbrace] (#1.west|-#1-#3-1.south west) -- node[left=2pt] {#4} (#1.west|-#1-#2-1.north west);
}

\newcommand*\mymatrixbracetop[4][]{
    \draw[mymatrixbrace] (#1.north-|#1-1-#2.north west) -- node[above=2pt] {#4} (#1.north-|#1-1-#3.north east);
}
\newcommand*\mymatrixbracebottom[4][]{
    \draw[mymatrixbrace] (#1.south-|#1-1-#3.north east)-- node[below=2pt] {#4} (#1.south-|#1-1-#2.north west);
}

\definecolor{mygreen}{rgb}{0.12, 0.3, 0.17}
\definecolor{myblue2}{rgb}{0.12, 0.3, 0.9}
\newcommand{\gf}{|[fill=mygreen!10]|}
\newcommand{\ggg}{|[fill=mygreen!20]|}
\newcommand{\gggg}{|[fill=mygreen!40]|}
\newcommand{\ggggg}{|[fill=mygreen!60]|}
\newcommand{\myteal}{|[fill=teal!30]|}

\newcommand{\mb}{|[fill=myblue2!20]|}
\newcommand{\mbb}{|[fill=myblue2!30]|}
\newcommand{\mbbb}{|[fill=myblue2!50]|}
\newcommand{\mbbbb}{|[fill=myblue2!70]|}

\newcommand{\cb}{|[fill=blue!25]|}
\newcommand{\cbb}{|[fill=blue!10]|}
\newcommand{\cbbb}{|[fill=blue!55]|}
\newcommand{\cbbbb}{|[fill=blue!40]|}

\usepackage{algorithm}
\usepackage{algorithmic}
\usepackage{appendix}

\RequirePackage{color}
\definecolor{greycolor}{cmyk}{0,0,0,0.8}
\definecolor{grey}{cmyk}{0,0,0,.1}
\definecolor{black}{cmyk}{0,0,0,1}
\definecolor{jmcolor}{cmyk}{0, 0.65, 0.90, 0}

\begin{document}

\title{HyPHEN: A Hybrid Packing Method and Its Optimizations for Homomorphic Encryption-based Neural Networks}

\author{
    Donghwan Kim\textsuperscript{*}, Jaiyoung Park\textsuperscript{*}, Jongmin Kim, Sangpyo Kim, and Jung Ho Ahn\\
    {Seoul National University, Seoul, South Korea}\\
    {\{{\it eastflame, jeff1273, jongmin.kim, vnb987, gajh}\}{\it @snu.ac.kr}}\\
}
\date{}
\maketitle

\begingroup\renewcommand\thefootnote{*}
\footnotetext{Equal contribution}
\endgroup

\begin{abstract}
Convolutional neural network (CNN) inference using fully homomorphic encryption (FHE) is a promising private inference (PI) solution due to the capability of FHE that enables offloading the whole computation process to the server while protecting the privacy of sensitive user data.
\reviewertwo{Prior FHE-based CNN (HCNN) work has demonstrated the feasibility of constructing deep neural network architectures such as ResNet using FHE.
Despite these advancements, HCNN still faces significant challenges in practicality due to the high computational and memory overhead. 
To overcome these limitations, we present HyPHEN, a deep HCNN construction that incorporates novel convolution algorithms (RAConv and CAConv), data packing methods (2D gap packing and PRCR scheme), and optimization techniques tailored to HCNN construction.}
Such enhancements enable HyPHEN to substantially reduce the memory footprint and the number of expensive homomorphic operations, such as ciphertext rotation and bootstrapping.
As a result, HyPHEN brings the latency of HCNN CIFAR-10 inference down to a practical level at 1.4 seconds (ResNet-20) and demonstrates HCNN ImageNet inference for the first time at 14.7 seconds (ResNet-18).

\end{abstract}

\section{Introduction}
\label{sec:introduction}

 Private inference (PI) has recently gained the spotlight in the machine-learning-as-a-service (MLaaS) domain, allowing cloud companies to comply with privacy regulations such as GDPR~\cite{eu_2016_gdpr} and HIPAA~\cite{act_1996_health}.
PI enables inference services at the cloud server while protecting both the privacy of the client and the intellectual properties of the service provider.
For example, by exploiting PI, hospitals can provide a private medical diagnosis of diseases, and security companies can provide private surveillance systems, each without accessing client’s sensitive data~\cite{kumar_2020_cryptflow, Bowditch_2020_surveillance}. 

\begin{figure}[t]
\centering
\includegraphics[width=0.95\columnwidth]{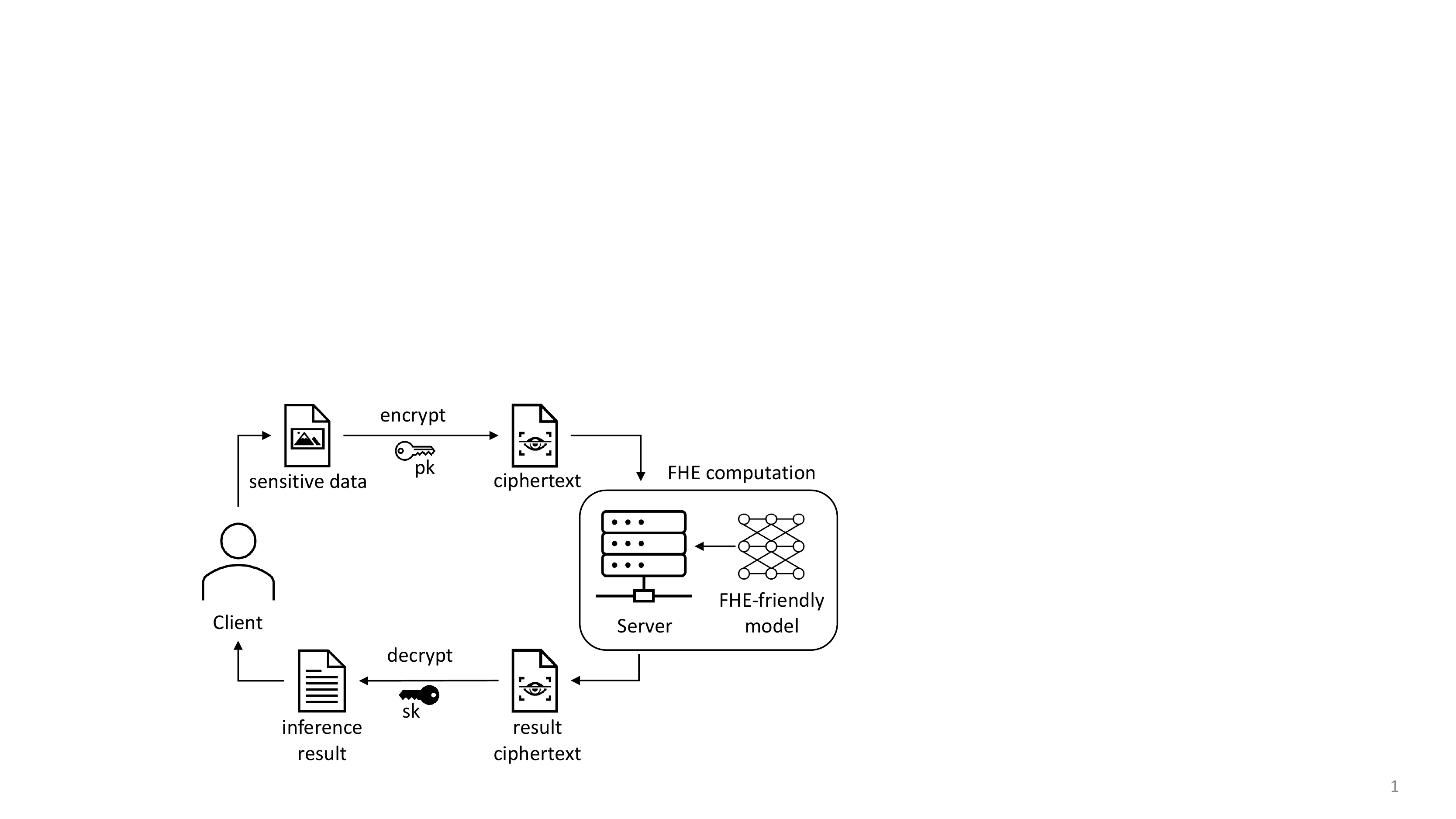} 
\caption{\reviewertwo{FHE-based private inference.} \label{schematic}}
\end{figure}

Fully homomorphic encryption (FHE)~\cite{gentry_2009_fully} is a cryptographic primitive that enables direct evaluation of a rich set of functions on encrypted data, making it especially suited for PI in terms of security and usability among other cryptographic candidates~\cite{yao_1982_protocols,costan_2016_intel}.
FHE-based PI solutions, \reviewertwo{illustrated in Figure~\ref{schematic}}, uniquely feature 1) full offloading of the computation process to the server, 2) succinct data communication requirement, and 3) non-disclosure of any information about the model except the inference result.
Such benefits, enabled by the unique capability of FHE that supports direct computation on ciphertexts (encrypted data), have driven researchers to investigate the FHE-based PI of convolutional neural networks (HCNN)~\cite{gilad_2016_cryptonets,brutzkus_2019_LoLa, dathathri_2020_eva,lee_2022_low, aharoni_2020_helayers}. 
In particular, this study explores HCNN construction with the CKKS FHE scheme~\cite{cheon_2017_homomorphic}, which offers higher throughput compared to the other FHE schemes and supports handling real and complex numbers.

Despite these benefits, FHE incurs high computational and memory overhead, which hinders the adoption of HCNN for real-world services.
Furthermore, computation with encrypted data in FHE exhibits distinct characteristics compared to their unencrypted counterpart, primarily in that the manipulation of data organization is extremely costly when encrypted.
Thus, to minimize the computational and memory overhead, HCNN requires an optimized convolution algorithm and a distinct data organization tailored to FHE circumstances.

Gazelle~\cite{juvekar_2018_gazelle}, a pioneering study in PI, provides an efficient convolutional algorithm that can be used in FHE to reduce the number of homomorphic operations compared to na\"ively adopting conventional convolution algorithms used for unencrypted CNN inference.
Gazelle avoids the high cost for the FHE evaluation of data rearrangement and nonlinear activation (e.g., ReLU) by combining the use of secure multi-party computation (MPC).
Still, Gazelle accumulates data elements dispersed in a ciphertext using FHE, which is also costly.
\cite{lee_2022_access} extends Gazelle's algorithm for end-to-end HCNN inference by introducing FHE-based data rearrangement and ReLU evaluation methods.
\cite{lee_2022_low} further enhances HCNN performance with a more dense data format, which minimizes the number of ciphertexts.
However, prior implementations take tens of minutes~\cite{lee_2022_low} to even hours~\cite{lee_2022_access} to perform a single HCNN inference for CIFAR-10 (ResNet-20) due to the high cost of accumulation, data rearrangement, and ReLU evaluation.
We can greatly reduce the cost for activation functions by utilizing activation functions proposed in AESPA~\cite{park_2022_aespa}; however, there still remains the cost for accumulation and data rearrangement.

We tackle this problem by combining multiple flexible data formats with a polymorphic FHE convolution algorithm that performs convolution tailored to each data format.
We identify that the high data rearrangement cost stems from the inconsistency in data format between the input and the output of a convolution layer.
Prior work utilizes a fixed data format throughout HCNN inference, incurring frequent data rearrangement using homomorphic rotation operations.
We instead propose using multiple flexible data formats by allowing replication of data elements inside a ciphertext.
The input and the output of a convolution layer can freely choose from multiple data formats to minimize the cost of data rearrangement.
To support different data formats, we also create polymorphic FHE convolution algorithms tailored to each data format.
The replication has a positive side effect of reducing the cost of data accumulation during convolution because fewer unique data elements need to be accumulated.

Furthermore, our data format enables scaling HCNN to real-world images and larger CNN models. In previous works~\cite{lee_2022_access,juvekar_2018_gazelle,lee_2022_low,brutzkus_2019_LoLa,dathathri_2020_eva}, memory expansion of plaintext weight has not received sufficient consideration. However, we identify that the memory footprint of weight plaintext increases significantly with image size and eventually brings a major bottleneck.
Thus, existing techniques do not scale to larger datasets such as ImageNet due to the substantial memory requirement for storing weights.
To address this challenge, we introduce a data formatting method that can effectively reduce the plaintext weight size. Our approach involves dividing an image into multiple image segments along the row direction, which ensures that the plaintext size correlates with the image segment size. Consequently, weight plaintexts become smaller and can be reused across these image segments during convolution, alleviating the memory footprint.

We evaluate the real-world scalability of HyPHEN, our HCNN framework combining the aforementioned solutions.
The GPU implementation of HyPHEN achieves 1.40 seconds for encrypted CIFAR-10 inference with the ResNet-20 model.
We also demonstrate for the first time end-to-end HCNN inference on the ImageNet dataset with the ResNet-18 model, achieving an execution time of 14.69 seconds.

The key contributions of the paper are as follows:

\begin{itemize}
    \item We propose a novel data format that enables streamlined data arrangement between consecutive convolutions.
    \item We devise optimized FHE convolution algorithms, which support various data formats with less computational complexity.
    \item We identify that huge memory footprint of HCNN deteriorates its performance and propose an efficient data format that can save hundreds of gigabytes of memory space with negligible overhead.
    \item We showcase the evaluation of various neural networks within practical execution times.
    We extend FHE-based PI to complex real-world data such as ImageNet, by demonstrating the performance with the ResNet-18 model.
\end{itemize}

\section{Background}
\label{subsec:Background}
\subsection{Fully Homomorphic Encryption (FHE)}
\label{subsec:FHE}
FHE is a set of public key encryption schemes that enable computation on encrypted data.
Among several popular FHE schemes, RNS-CKKS~\cite{cheon_18_rns} has been broadly adopted in the PI domain as it supports fixed-point numbers and \emph{slot batching}.
A \emph{\ptxt} in RNS-CKKS is an unencrypted polynomial in a cyclotomic polynomial ring $\mathbb{Z}[X]/(X^N+1)$ for a power-of-two degree $N$, whose typical values are $2^{15}$--$2^{17}$.
A plaintext can be encrypted into a \emph{ciphertext}, which is a pair of polynomials hiding the plaintext using an additional random polynomial and an obfuscating error term.
As a vector containing $N/2$ real (or complex) numbers, referred to as \emph{message}, can be mapped to a plaintext, we can simultaneously operate on $N/2$ numbers by performing homomorphic operations on plaintexts and ciphertexts. This is called slot batching because the position inside this vector message is referred to as \emph{slot}.
It is also viable to batch a message with a shorter (power-of-two) length.

The RNS-CKKS scheme supports slot-wise additive (AddPt/AddCt) and multiplicative (MulPt/MulCt) operations, some of which are shown in Table~\ref{tb:Benchmark} with their respective execution time.
AddCt/MulCt, for instance, receives two ciphertexts as input and returns a ciphertext corresponding to a vector message which is approximately equal to the element-wise addition/multiplication result between the two vector messages contained in the input ciphertexts.
AddPt/MulPt does a similar job except that it receives a plaintext and a ciphertext as input and that it involves less computation than AddCt/MulCt.
In particular, MulPt takes 34$\times$ less execution time compared to MulCt.

A fundamental limitation of HE is that the number of sequential multiplication with a ciphertext is limited.
A ciphertext or plaintext is associated with a (multiplicative) level $\ell$, where $0 \leq \ell \leq L$ for the max level $L$.
The size of a ciphertext or plaintext and the complexity of homomorphic operations increases with the level.
After MulCt or MulPt, Rescale should be performed to the output ciphertext, which reduces the amplified error in the ciphertext due to multiplication.
Rescale reduces the level by one and the level cannot go below zero.
Also, to perform operations with operands having different levels, Rescale needs to be performed to adjust the operands' levels to the lowest among their levels.

To overcome the limitation, bootstrapping (Boot in Table~\ref{tb:Benchmark}) is required, which is a unique operation for increasing the level of a ciphertext.
With bootstrapping, a ciphertext's level can be increased up to $L^\prime$, which is smaller than the max level $L$ due to the levels bootstrapping consumes for its computation.
Therefore, in practice, we can only utilize $L^\prime$ levels for other operations.
As can be observed from Table~\ref{tb:Benchmark}, bootstrapping is orders of magnitude more expensive than basic operations.
Therefore, it is crucial to suppress level consumption and perform bootstrapping as few as possible.

Another limitation of HE is that, when using slot batching, it is difficult to arbitrarily change the data order among slots.
The only available option is cyclically shifting the slots, which we refer to as rotation.
CRot is a homomorphic operation to rotate a ciphertext to the left by a given amount.
When computation between data elements in different slots is required, CRot is performed to adjust the positions.
As such a computational pattern is extremely common, CRot is one of the most frequently performed operation in RNS-CKKS applications.
Also, as CRot is a relatively expensive operation among basic operations (see Table~\ref{tb:Benchmark}), CRot accounts for a large portion of computational overhead.
There is also a rotation operation for plaintexts, PRot, but it is rarely performed and is computationally cheap.

The data organization among the slots is the paramount concern in RNS-CKKS because it determines the number of CRot and bootstrapping operations.
The cost of other operations is much less sensitive to the data organization.
To reduce the number of bootstrapping, it is advisable to pack as many data elements as possible into a ciphertext to reduce the number of ciphertexts to bootstrap.
However, such a dense packing may result in increased amounts of rotations because putting some data elements instead into another ciphertext would remove the need for rotation.
As an extreme example, if we pack only one element per ciphertext, we can eliminate all rotation operations; however, this would incur excessively high cost for bootstrapping and other operations.
Therefore, to deliver high performance for RNS-CKKS applications, we need to devise an application-specific data organization that strikes a balance between the two objectives, minimizing computation among different slots and maximizing the slot usage.

\setlength{\tabcolsep}{2pt}
\renewcommand{\arraystretch}{1}
\begin{table}[t]
\centering
    \caption{Benchmark of homomorphic operations averaged over 100 iterations on CPU (64 threads). Pt and Ct postfixes each represents \ctxt-\ptxt and \ctxt-\ctxt operation, respectively. PRot and CRot express the rotation of plaintext and ciphertext, respectively. The experimental setup is detailed in Section~\ref{subsec:Benchmarks}. \label{tb:Benchmark}}
\Scale[0.9]{
{\small
\begin{tabular}{ccccccccc}
\toprule[1.0pt]
\textbf{Operation} & AddPt & AddCt & MulPt & MulCt & Rescale & 
PRot & CRot & Boot    \\
\midrule[0.4pt]
\textbf{Time (ms)}  & 0.169   & 0.202   & 0.506  & 17.3  & 3.90  & 0.102   & 15.5  & 2160   \\
\bottomrule[1.0pt]
\end{tabular}}
}
\end{table}
\setlength{\tabcolsep}{6pt}

\subsection{Convolution on HE}
\label{sec:convolution_background}

\begin{figure*}
    \centering
    \subfloat[Unencrypted convolution for $s, pad=1$]{\label{fig:convolution}
        \centering
        \scalebox{0.65}{
        \begin{tikzpicture}[mymatrixenv]
    \matrix(mtr)[mystyle2, font=\large]{
    $a_{1}$&$a_{2}$&$a_{3}$&$a_{4}$\\
    $a_{5}$&$a_{6}$&$a_{7}$&$a_{8}$\\
    $a_{9}$&$a_{10}$&$a_{11}$&$a_{12}$\\
    $a_{13}$&$a_{14}$&$a_{15}$&$a_{16}$\\
    };
    \draw[thick, black] (mtr-1-1.north west) rectangle (mtr-4-4.south east);

    \begin{scope}[every node/.append style={scale=\myscale,transform
    shape},very thick]
        \mymatrixbraceright[mtr]{1}{4}{$w_i$}
        \mymatrixbracetop[mtr]{1}{4}{$w_i$}
    \end{scope} 
    \node[right = 0em of mtr, scale=1.2] (str) {$*$};

    \matrix(K)[style3, right=1.5em of str, font=\large]{
    $k_{1}$&$k_{2}$&$k_{3}$\\
    $k_{4}$&$k_{5}$&$k_{6}$\\
    $k_{7}$&$k_{8}$&$k_{9}$\\
    };
    \draw[thick, black] (K-1-1.north west) rectangle (K-3-3.south east);

    \begin{scope}[every node/.append style={scale=\myscale,transform
    shape},very thick]
        \mymatrixbraceright[K]{1}{3}{$f$}
        \mymatrixbracetop[K]{1}{3}{$f$}
    \end{scope} 
	\node [right = 0em of K, scale=1.2] (eq) {$=$};

    \matrix(ret)[style2, right=1.5em of eq, font=\large]{
    $c_{1}$&$c_{2}$&$c_{3}$&$c_{4}$\\
    $c_{5}$&$c_{6}$&$c_{7}$&$c_{8}$\\
    $c_{9}$&$c_{10}$&$c_{11}$&$c_{12}$\\
    $c_{13}$&$c_{14}$&$c_{15}$&$c_{16}$\\
    };
    \draw[thick, black] (ret-1-1.north west) rectangle (ret-4-4.south east);

    \begin{scope}[every node/.append style={scale=\myscale,transform
    shape},very thick]
        \mymatrixbraceright[ret]{1}{4}{$w_o$}
        \mymatrixbracetop[ret]{1}{4}{$w_o$}
    \end{scope} 
\end{tikzpicture}}}
        \hfill
    \centering
    \subfloat[Slide$_f$ for $f=3$]{\label{fig:siso_rot}
        \centering
        \scalebox{0.65}{
        \begin{tikzpicture}[mymatrixenv]
    \matrix(m1)[mystyle2,font=\large]{
    $a_{1}$&$a_{2}$&$a_{3}$&$a_{4}$\\
    $a_{5}$&$a_{6}$&$a_{7}$&$a_{8}$\\
    $a_{9}$&$a_{10}$&$a_{11}$&$a_{12}$\\
    $a_{13}$&$a_{14}$&$a_{15}$&$a_{16}$\\
    };
    
    \node[right = 0em of m1] (str1) {\Large $\xrightarrow{(f^2-1)\, Rots}$};

    \draw[thick, black] (m1-1-1.north west) rectangle (m1-4-4.south east);

    \matrix(f1)[mystyle2, right= -0.4em of str1,font=\large]{
    \myhash&\myhash&\myhash&\myhash\\
    \myhash&$a_{1}$&$a_{2}$&$a_{3}$\\
    \myhash&$a_{5}$&$a_{6}$&$a_{7}$\\
    \myhash&$a_{9}$&$a_{10}$&$a_{11}$\\
    };
    \draw[thick, black] (f1-1-1.north west) rectangle (f1-4-4.south east);
    \node [right = 0em of f1-4-4.south east, scale=1.2] (eq1) {\Large $,$};

    \matrix(f2)[mystyle2,right= 1em of f1, font=\large]{
    \myhash&\myhash&\myhash&\myhash\\
    $a_{1}$&$a_{2}$&$a_{3}$&$a_{4}$\\
    $a_{5}$&$a_{6}$&$a_{7}$&$a_{8}$\\
    $a_{9}$&$a_{10}$&$a_{11}$&$a_{12}$\\
    };
    \draw[thick, black] (f2-1-1.north west) rectangle (f2-4-4.south east);
    \node [right = -0.4em of f2, scale=1.2] (eq2) {$\cdots$};
\end{tikzpicture}}}
        \hfill
\subfloat[MulFilter\&Sum$_f$ for $s, pad=1$]{\label{fig:siso_stride1}
	    \centering
     \scalebox{0.65}{
		\begin{tikzpicture}[mymatrixenv]
    \matrix(m1)[mystyle2,font=\large]{
    \myhash&\myhash&\myhash&\myhash\\
    \myhash&$a_{1}$&$a_{2}$&$a_{3}$\\
    \myhash&$a_{5}$&$a_{6}$&$a_{7}$\\
    \myhash&$a_{9}$&$a_{10}$&$a_{11}$\\
    };
    \node[right = -0.4em of m1, scale=1.2] (str1) {$\odot$};
    \draw[thick, black] (m1-1-1.north west) rectangle (m1-4-4.south east);

    \matrix(f1)[style3, right= -0.4em of str1,font=\large]{
    \myzero&\myzero&\myzero&\myzero\\
    \myzero&$k_{1}$&$k_{1}$&$k_{1}$\\
    \myzero&$k_{1}$&$k_{1}$&$k_{1}$\\
    \myzero&$k_{1}$&$k_{1}$&$k_{1}$\\
    };
    \node [right =-0.4em of f1, scale=1.2] (eq1) {$\oplus$};
    \draw[thick, black] (f1-1-1.north west) rectangle (f1-4-4.south east);

    \matrix(m2)[mystyle2,right= -0.4em of eq1, font=\large]{
    \myhash&\myhash&\myhash&\myhash\\
    $a_{1}$&$a_{2}$&$a_{3}$&$a_{4}$\\
    $a_{5}$&$a_{6}$&$a_{7}$&$a_{8}$\\
    $a_{9}$&$a_{10}$&$a_{11}$&$a_{12}$\\
    };
    \node[right = -0.4em of m2, scale=1.2] (str2) {$\odot$};
    \draw[thick, black] (m2-1-1.north west) rectangle (m2-4-4.south east);

    \matrix(f2)[style3, right= -0.4em of str2,font=\large]{
    \myzero&\myzero&\myzero&\myzero\\
    $k_{2}$&$k_{2}$&$k_{2}$&$k_{2}$\\
    $k_{2}$&$k_{2}$&$k_{2}$&$k_{2}$\\
    $k_{2}$&$k_{2}$&$k_{2}$&$k_{2}$\\
    };
    \node [right = -0.4em of f2, scale=1.2] (eq2) {$\cdots=$};
    \draw[thick, black] (f2-1-1.north west) rectangle (f2-4-4.south east);
    
    \matrix(ret)[style2, right= -0.4em of eq2,font=\large]{
    $c_{1}$&$c_{2}$&$c_{3}$&$c_{4}$\\
    $c_{5}$&$c_{6}$&$c_{7}$&$c_{8}$\\
    $c_{9}$&$c_{10}$&$c_{11}$&$c_{12}$\\
    $c_{13}$&$c_{14}$&$c_{15}$&$c_{16}$\\
    };
    \draw[thick, black] (ret-1-1.north west) rectangle (ret-4-4.south east);

\end{tikzpicture}}}

	    \centering
\subfloat[MulFilter\&Sum$_f$ for $s=2, pad=1$]{\label{fig:siso_stride2}
	    \scalebox{0.65}{
		\begin{tikzpicture}[mymatrixenv]
    \matrix(m1)[mystyle2, font=\large]{
    \myhash&\myhash&\myhash&\myhash\\
    \myhash&$a_{1}$&$a_{2}$&$a_{3}$\\
    \myhash&$a_{5}$&$a_{6}$&$a_{7}$\\
    \myhash&$a_{9}$&$a_{10}$&$a_{11}$\\
    };
\node[right = -0.4em of m1, scale=1.2] (str1) {$\odot$};
\draw[thick, black] (m1-1-1.north west) rectangle (m1-4-4.south east);

\matrix(f1)[style3, right=-0.4em of str1, font=\large]{
    \myzero&\myzero&\myzero&\myzero\\
    \myzero&\myzero&\myzero&\myzero\\
    \myzero&\myzero&$k_{1}$&\myzero\\
    \myzero&\myzero&\myzero&\myzero\\
};
\draw[thick, black] (f1-1-1.north west) rectangle (f1-4-4.south east);
\node [right = -0.4em of f1, scale=1.2] (eq1) {$\oplus$};

\matrix(m2)[mystyle2,right= -0.4em of eq1, font=\large]{
    \myhash&\myhash&\myhash&\myhash\\
    $a_{1}$&$a_{2}$&$a_{3}$&$a_{4}$\\
    $a_{5}$&$a_{6}$&$a_{7}$&$a_{8}$\\
    $a_{9}$&$a_{10}$&$a_{11}$&$a_{12}$\\
};
\draw[thick, black] (m2-1-1.north west) rectangle (m2-4-4.south east);
\node[right = -0.4em of m2, scale=1.2] (str2) {$\odot$};

\matrix(f2)[style3, right= -0.4em of str2, font=\large]{
\myzero&\myzero&\myzero&\myzero\\
\myzero&\myzero&\myzero&\myzero\\
$k_{2}$&\myzero&$k_{2}$&\myzero\\
\myzero&\myzero&\myzero&\myzero\\
};
\node [right = -0.4em of f2, scale=1.2] (eq2) {$\cdots=$};
\draw[thick, black] (f2-1-1.north west) rectangle (f2-4-4.south east);

\matrix(ret)[style2, right= 0em of eq2, font=\large]{
$c_{1}$&\myzero&$c_{2}$&\myzero\\
\myzero&\myzero&\myzero&\myzero\\
$c_{3}$&\myzero&$c_{4}$&\myzero\\
\myzero&\myzero&\myzero&\myzero\\
};
\draw[thick, black] (ret-1-1.north west) rectangle (ret-4-4.south east);

\begin{scope}[every node/.append style={scale=\myscale,transform
shape},very thick]
    \mymatrixbraceright[ret]{1}{2}{$g$}
    \mymatrixbracetop[ret]{1}{2}{$g$}
\end{scope} 
    
\end{tikzpicture}}}
		
    \caption{Single-input and single-output channel convolution (SISO)\reviewertwo{~\cite{juvekar_2018_gazelle}}. Image ciphertexts and filter plaintexts are illustrated as 2D matrices, but are stored in a 1D manner with each matrix row concatenated. $\odot$ symbolizes MulPt and $\oplus$ symbolizes AddCt. }
    \label{fig:siso}
\end{figure*}
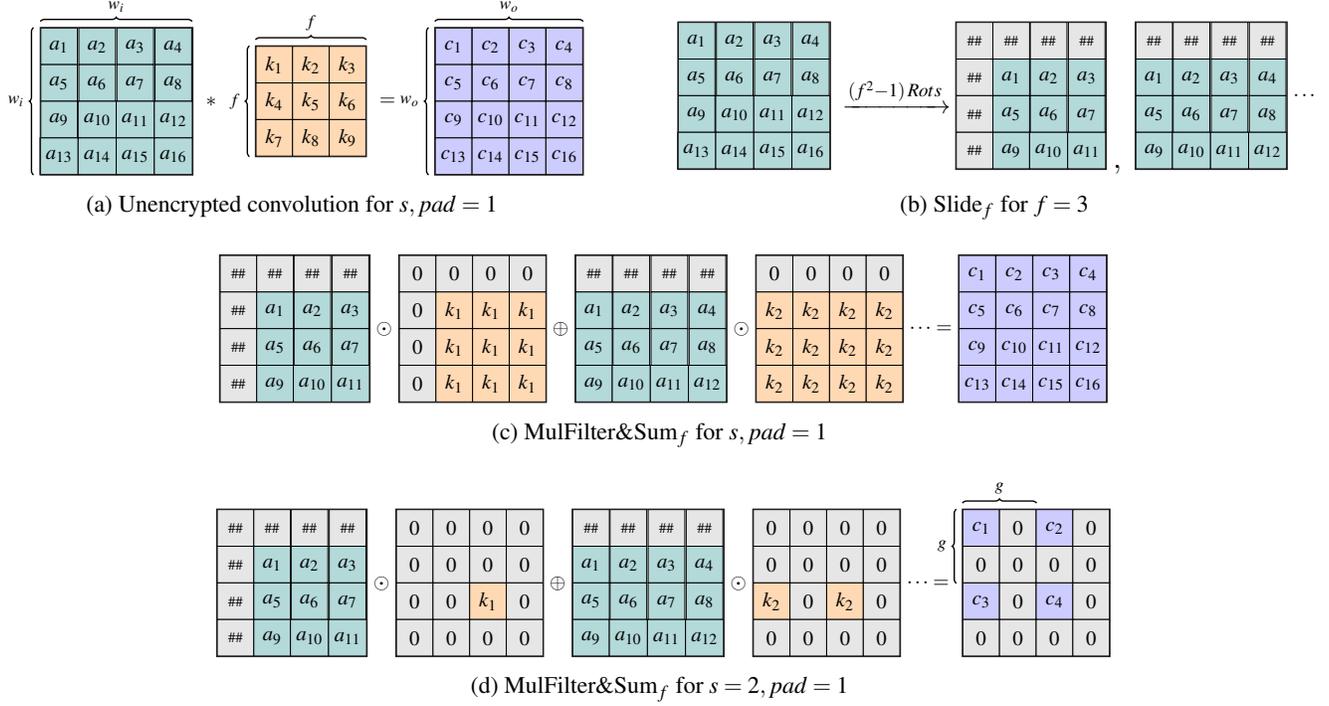

We first describe the notations for the baseline CNN network.
For the simplicity of notation, we assume images and filters are square.
We represent input and output of a single convolutional layer with tuples $\{$channel, image width$\}$ as $\{c_i,w_i\}$ and $\{c_o,w_o\}$, respectively.
Parameters for a convolution are represented with tuples $\{$output channel, input channel, filter width$\}$ denoted as $\{c_o,c_i, f\}$, stride $s$, and padding $pad$.

\begin{algorithm}[t]
\caption{\reviewertwo{Single-input, single-output channel convolution (SISO)} \label{alg:SISO}}
\textbf{Input} $ct_{i}$: input ciphertext, $W$: plaintext filter\\
\textbf{Output} $ct_{o}$: output ciphertext
\begin{algorithmic}[1]
    \STATE {// SubRoutine: \textbf{Slide$_f$}($ct_i$) }
    \FOR {$j_1= 0 ,\ldots,f-1 $}
        \FOR{$j_2= 0 ,\ldots,f-1 $}
            \STATE $r \leftarrow w_i (j_1-\frac{f-1}{2}) + (j_2 - \frac{f-1}{2})$
            \STATE $ct'[j_1,j_2]$ $\leftarrow$ CRot($ct_{i}; r$)
        \ENDFOR
    \ENDFOR
    \STATE {// SubRoutine: \textbf{MulFilter}\&\textbf{Sum$_f$($ct_i,W$)}}
        \FOR {$j_1= 0 ,\ldots,f-1 $}
            \FOR{$j_2= 0 ,\ldots,f-1 $}
                \STATE $ct_o$ $\mathrel{+}=$ MulPt($ct'[j_1,j_2], W[j_1,j_2]$) 
            \ENDFOR
        \ENDFOR
    \STATE \textbf{return} $ct_{o}$
\end{algorithmic}
\end{algorithm}

\begin{figure*}[t]
		\centering 
	    \scalebox{0.5}{\includegraphics{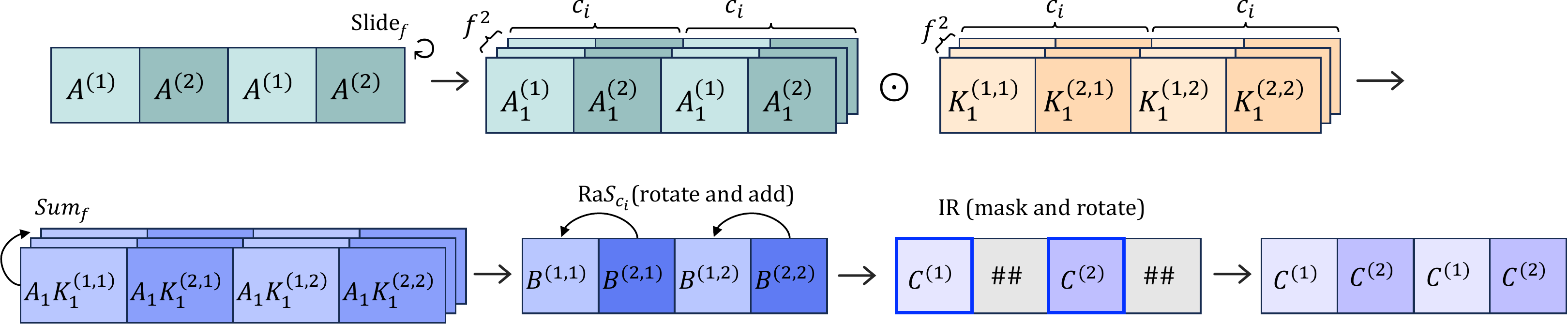}}\\\vfill \hfill	 
     \vspace{-0.1in}
	\caption{Convolution procedure in \cite{lee_2022_low} when $c_i,c_o=2$. A single superscript denotes the channel and a superscript pair denotes (input channel, output channel). $A^{(m)}$ represents the $m$-th channel of the input image contiguously placed in the slots and $K^{(m,n)}$ is the corresponding filter deployed in plaintexts. We simplify the notation of the intermediate SISO result $\sum_{i=1}^{f^2} {CRot(A^{(m)};r(i))K_i^{(m,n)}}$ as ${B}^{(m,n)}$. $C^{(n)}$ represents the $n$-th channel of the output image. \label{fig:lee_mcconv}}
\end{figure*}

\textbf{Single-input, single-output channel convolution (SISO):} 
Gazelle~\cite{juvekar_2018_gazelle} proposes an efficient convolution algorithm on HE, referred to as SISO.
Figure~\ref{fig:siso} illustrates SISO with $s$ equal to 1 and 2, where an input ciphertext contains $w^2_i$ pixels.
Although we represent the data format with two dimensions as $(H, W)$ in this example, pixels are stored in the slots in a flattened row-major order regardless of the number of dimensions.
\reviewertwo{$f^2$ plaintexts each storing distinct filter elements need to be prepared for SISO.
Each slot of the $i$-th plaintext ($0\leq i < f^2 $) holds a filter element ($k_i$) or zero (0),
depending on whether $k_i$ participates in the computation of the output pixel at the same slot.}
\reviewertwo{SISO operation proceeds as follows (see Algorithm~\ref{alg:SISO}):}
\begin{enumerate}
\item \reviewertwo{\emph{Slide$_f$} rotates an input image placed in encrypted ciphertext with different rotation amounts for each of the $f^2$ plaintexts as shown in Figure~\ref{fig:siso_rot}.}
\item \reviewertwo{\emph{MulFilter\&Sum$_f$} multiplies each rotated input by a filter plaintext and accumulates each of the $f^2$ multiplied results to obtain the output as depicted in Figure~\ref{fig:siso_stride1}.}
\end{enumerate}
As an input image tuple in CNN usually has multiple channels, multiple ciphertexts are often required per convolutional layer when using SISO.
We denote the number of input and output ciphertexts per convolutional layer as $n_i$ and $n_o$, respectively.

\textbf{Convolution and data format:}
When the image size ($w_i^2$) is smaller than $N/2$, we can reduce $n_i$ and $n_o$ by batching multiple channels into a single ciphertext.
Gazelle also proposes a \emph{channel-aligned} batching method where the data format of the input and output ciphertexts can be regarded as a flattened ${(C, H, W)}$ 3D vector. 
\cite{lee_2022_access} follows it and utilizes the same ${(C, H, W)}$ format to implement an end-to-end CNN inference using FHE.
By following steps similar to SISO, this format enables performing convolutions on multiple channels simultaneously.
Moreover, \cite{lee_2022_low} introduces input repetition to further enhance parallelism, where the data format can be represented as ${(R,C,H,W)}$ and the input tuple is repeated $|R|$ times to fill all the slots of a ciphertext.
In Figure~\ref{fig:lee_mcconv}, we present an example when $\{c_o, c_i, w_i\} = \{2, 2, 32\}$ and the number of slots ($N/2$) is 4,096. 
The input ciphertext ($n_i = 1$) has two channels ($A^{(1)}, A^{(2)}$) repeated twice.
$f^2$ filter plaintexts are prepared, where each plaintext holds filter elements with $|C|$ input channels for $|R|$ output channels.
Convolution on this ciphertext~\cite{lee_2022_low} can be described as follows:
\begin{enumerate}
\item SISO: With a single input ciphertext, SISO can be performed to get convolution results for $|C|$ input channels and $|R|$ output channels simultaneously. The resulting ciphertext contains $|R||C|$ intermediate convolution outputs $B^{(x,y)}$ ($1\leq x\leq |C|$, $1\leq y \leq |R|$).
\item RaS$_{c_{i}}$: To obtain the result for the $y$-th output channel, accumulation of $|C|$ intermediate results are performed ($\sum_{x=1}^{|C|}B^{(x,y)}$) by computing \emph{rotate and sum} (RaS), which requires $\log|C|$ rotations.
\item IR: To proceed with further convolutions, the data format should be rearranged to match the next layer's data format by masking data elements and rotating them to reposition them, which we refer to as \emph{image realigning} (IR).
\end{enumerate}
Throughout this paper, we refer to this convolution that takes an ${(R,C,H,W)}$ ciphertext as \emph{$Conv_{lc}$}.

\textbf{Gap and multiplexed packing:} Strided convolution ($s>1$) using SISO generates a \emph{gap} (denoted as $g$) between valid values (see Figure~\ref{fig:siso_stride2}).
When using a \ctxt with a gap, slot underutilization degrades the throughput.
While interactive protocols such as Gazelle~\cite{juvekar_2018_gazelle} removes the gap by a client-aided re-encryption process, FHE demands heavy masking (MulPt) and rotation (CRot) operations to remove the gap,
which incurs additional computation and level consumption.
\cite{lee_2022_low} proposes \emph{multiplexed packing} (MP) method on top of $Conv_{lc}$ (\emph{$MP\text{-}Conv_{lc}$}) to remedy the slot underutilization.

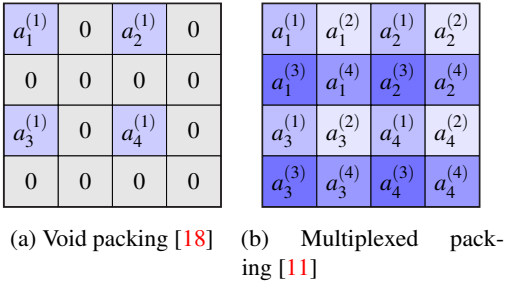
\begin{figure}[t]
    \hfil
        \subfloat[Void packing~\cite{lee_2021_precise}]{\label{fig:packing_gap}
        
            \centering\scalebox{0.9}{
    	        \begin{tikzpicture}[mymatrixenv]
    \matrix(m1)[style2_mid]{
    $a_{1}^{(1)}$&\myzero&$a_{2}^{(1)}$&\myzero\\
    \myzero&\myzero&\myzero&\myzero\\
    $a_{3}^{(1)}$&\myzero&$a_{4}^{(1)}$&\myzero\\
    \myzero&\myzero&\myzero&\myzero\\
    };
\draw[semithick, black] (m1-1-1.north west) rectangle (m1-4-4.south east);
\end{tikzpicture}}
    	        }
        \subfloat[Multiplexed packing~\cite{lee_2022_low}]{\label{fig:packing_mpx}
            \centering\scalebox{0.9}{
    	        \begin{tikzpicture}[mymatrixenv]
    \matrix(m1)[style2_mid]{
    \cb$a_1^{(1)}$&\cbb$a_{1}^{(2)}$&\cb$a_{2}^{(1)}$&\cbb$a_{2}^{(2)}$\\
    \cbbb$a_{1}^{(3)}$&\cbbbb$a_{1}^{(4)}$&\cbbb$a_{2}^{(3)}$&\cbbbb$a_{2}^{(4)}$\\
    \cb$a_{3}^{(1)}$&\cbb$a_{3}^{(2)}$&\cb$a_{4}^{(1)}$&\cbb$a_{4}^{(2)}$\\
    \cbbb$a_{3}^{(3)}$&\cbbbb$a_{3}^{(4)}$&\cbbb$a_{4}^{(3)}$&\cbbbb$a_{4}^{(4)}$\\
    };
\draw[semithick, black] (m1-1-1.north west) rectangle (m1-4-4.south east);
\end{tikzpicture}}
    	        }
             \hfil
	\caption{Previous gap packing methods to fill gap induced by downsampling layers. $a_i^{(j)}$ denotes the $i$-th element in the $j$-th channel of an image. \label{fig:packing}}
\end{figure}

$MP\text{-}Conv_{lc}$ adds \emph{repacking process} to the IR process of $Conv_{lc}$, which fills the gap with different channels (see Figure~\ref{fig:packing_mpx}).
$MP\text{-}Conv_{lc}$ uses the data format of $(R,C_a,H,W,C_g)$, adding mutiplexed channel at the innermost dimension $C_g$.
The convolution process on an encrypted image with $|C_g|=4$ is depicted in Figure~\ref{fig:hybrid_IAConv}.
After applying SISO and RaS on the outer channel dimension $C_a$, additional RaS is performed to accumulate channels in the inner dimension $C_g$ located inside the gap.
Multiplexed packing requires a more complex IR process which includes filling the gap with multiple channels from the output tensor.
For further details of $MP\text{-}Conv_{lc}$, we refer the readers to Section 4 of \cite{lee_2022_low}.

\begin{figure}[t]
\centering
\includegraphics[width=0.85\columnwidth]{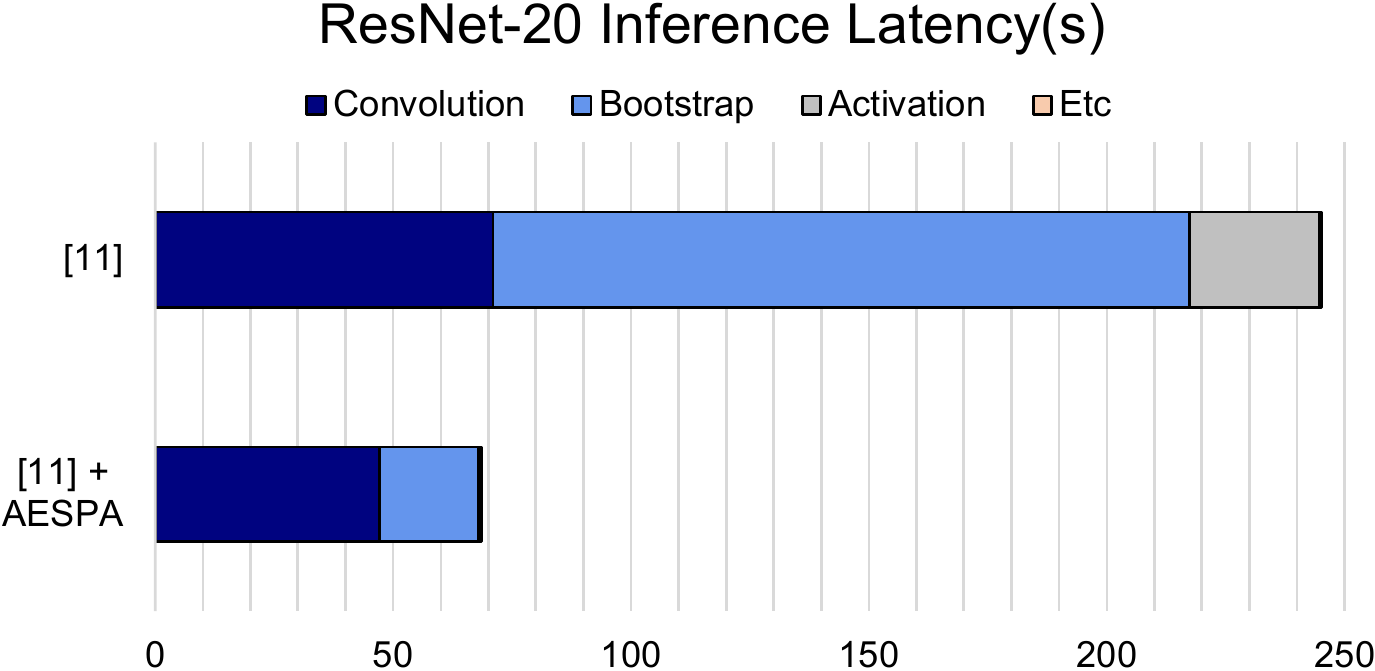}
\caption{ResNet-20 inference latency and their breakdown for ~\cite{lee_2022_low} and the adoption of low-degree polynomial~\cite{park_2022_aespa}. Two experiments are conducted under paramter Set$_{lc}$ and Set$_hyp$ from ~\ref{tb:parameter set}, respectively. \label{fig:RotRuntime}}
\end{figure}

\subsection{Activation Function on HE}
Nonlinear activation functions, such as ReLU, cannot be used directly in HCNN. They must be replaced by polynomial functions approximating them because FHE only supports additive and multiplicative operations.
Maintaining a low polynomial approximation error across a wide range is crucial to preserve the accuracy of a CNN model.
\cite{lee_2022_access} approximates ReLU with a composition of 15-, 15-, and 27-degree polynomials, which keeps L1 norm of approximation error lower than $2^{-13}$ in range [-50, 50].
This approach has a benefit that pretrained CNN models can be directly used without modification.
However, the evaluation of high-degree polynomials imposes a significant runtime overhead; a series of Rescale during the evaluation incurs a lot of level consumption, resulting in an increased number of bootstrapping.
One may attempt to mitigate this overhead by using a CKKS parameter with more levels (i.e., higher $L^\prime$); however, such a parameter set has an extremely large memory footprint as presented in Table~\ref{tb:parameter set}, and thus each operation becomes much more expensive.

Another approach is to retrain neural networks with low-degree polynomial activation functions as in 
\cite{ishiyama_2020_highly,Chabanne_2017_privacy,obla_2020_effective,hesamifard_2019_deep,thaine_2019_efficient}.
By retraining, the operational cost drops significantly.
Recently, AESPA~\cite{park_2022_aespa} has shown that CNNs trained with low-degree polynomials can achieve equivalent accuracy to the original ReLU-based networks across various CNN architectures and image datasets.
AESPA replaces ReLU and batch normalization (BN) with the composition of orthogonal basis polynomials and basis-wise BN.
During inference, vertical layer fusion transforms the composition into a simple square function, drastically reducing the runtime of activation.

Due to the use of high-degree polynomials, the primary performance bottleneck of prior work~\cite{lee_2022_low} stems from bootstrapping operations.
In contrast, our analysis reveals that, if we adopt AESPA, the portion of bootstrapping in the entire HCNN inference time becomes small and convolution operations dominate the execution time (see Figure~\ref{fig:RotRuntime}).
We also analyze that rotation operations (CRot) account for the most of computation in convolution.
Therefore, in this work, we focus on enhancing the performance of convolution by proposing convolution algorithms and packing methods that effectively mitigates the substantial rotation overhead in HCNN inference.

\begin{figure*}[t]
		\centering
	\subfloat[CAConv when $|C_a|=c_i=c_o=2$]{\label{fig:alternate1_nostride}
	    \scalebox{0.5}{
	    \includegraphics{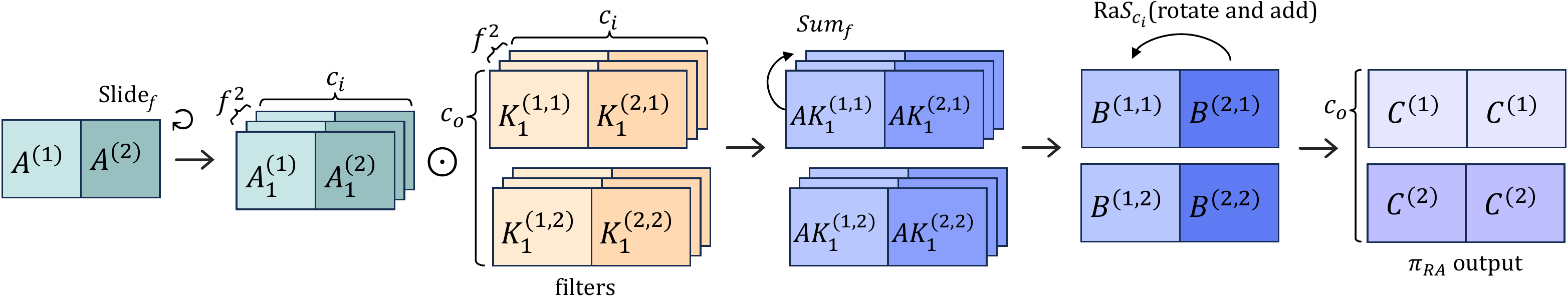}}}
            \hfill 
	\centering
	\subfloat[$RAConv_{Naive}$ when $|R_a|=c_i=c_o=2$.  ]{\label{fig:alternate2_nostride}
	\scalebox{0.5}{
    \includegraphics{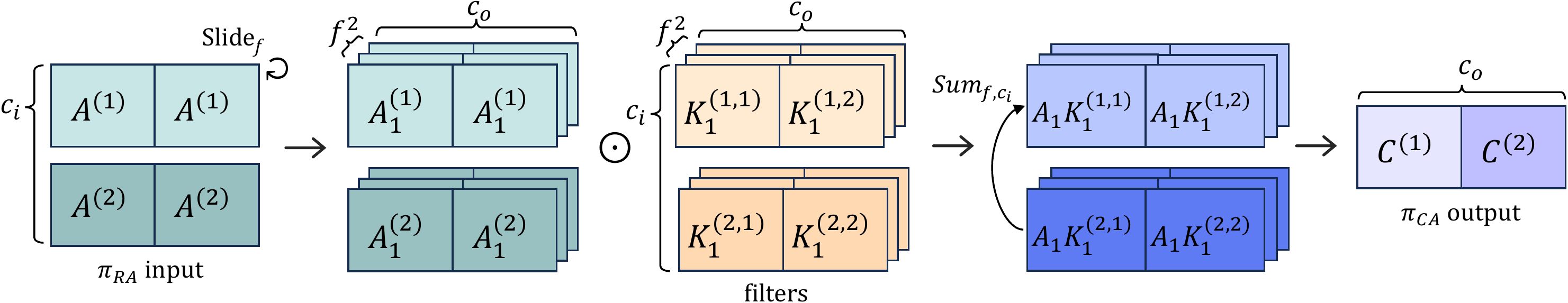}}}
	    \hfill
	    \centering
	\subfloat[$RAConv_{Reorder}$ when $|R_a|=c_i=c_o=2$.]{\label{fig:alternate2_nostride_lazy}
	    \scalebox{0.5}{
	    \includegraphics{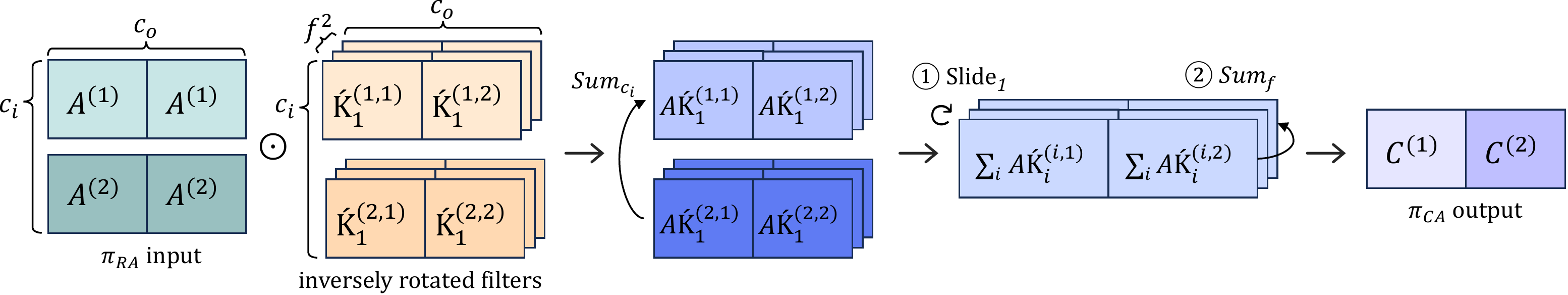}}}
	\hfill    
	\caption{CAConv and two variants of RAConv. A single superscript denotes the channel and a superscript pair denotes (input channel, output channel). We simplify the notation of $M^{(a)}K^{(a,b)}$ as ${MK}^{(a,b)}$.\label{fig:mc-siso}}
\end{figure*}

\section{HyPHEN Construction}
\label{sec:Method}

We introduce HyPHEN, our HCNN solution that focuses on reducing the memory footprint and the number of resource-intensive homomorphic operations, including rotation and bootstrapping.
We propose convolution algorithms and data formats that can streamline data arrangement between consecutive convolutions.
Our method consists of two data formats: channel aligned (CA) and replication aligned (RA) formats, which we denote as $\pi_{CA}$ and $\pi_{RA}$.
Both formats can be formally described as $\pi_{CA} = \{C_a,H,W,R_g,C_g\}$, $\pi_{RA} = \{R_a,H,W,C_g,R_g\}$.
In the outermost dimension, $\pi_{CA}$ aligns images with different channels ($C_a$) similar to \cite{juvekar_2018_gazelle, lee_2021_precise}, whereas $\pi_{RA}$ aligns with the replications ($R_a$) of images. 
We introduce two distinct convolutions for the two data formats in the following section.

\subsection{Convolution Algorithms of HyPHEN}
\label{sec:Convolution Algorithms}
We devise two convolution algorithms that start with input ciphertexts in the format of $\pi_{CA}$ and $\pi_{RA}$, respectively (see Figure~\ref{fig:mc-siso}).
For the simplicity of illustration, we assume that the size of the last two dimensions is one, implying that no gap exists, in the figure; gap packing is handled by additional computation during RaS and IR, which will be discussed in Section~\ref{subsec:2d_gap_packing}.
We focus on how the outermost $C_a$ ($R_a$) dimension is handled for $\pi_{CA}$ ($\pi_{RA}$).

\textbf{Channel-aligned convolution (CAConv)} is designed for the $\pi_{CA}$ data format, resembling $MP\text{-}Conv_{lc}$ but without requiring the IR process. 
Recall that, in $MP\text{-}Conv_{lc}$ targeting the data format $\{R,C_a,H,W,C_g\}$, $|C_a|$ ($|C_g| = 1$ for simplicity) intermediate SISO results are accumulated into the first channel's position and the rest of the slots contains meaningless values (\#\# in Figure~\ref{fig:lee_mcconv}). 
Subsequent IR rearranges the ciphertext for the next convolution.
In contrast, by relocating the input repetition to inside the gap ($\pi_{CA} = \{C_a,H,W,R_g,C_g\}$), we can position $C_a$ as the outermost dimension.
Executing RaS on this ciphertext automatically leads to the replication of the output result because homomorphic rotation is cyclic.
The resulting ciphertexts follow the data format of $\pi_{RA}$ as depicted in Figure~\ref{fig:alternate1_nostride}.
By using CAConv, our image realignment does not require any rotation at the cost of an increase in the number of output ciphertext by $c_o$.

\textbf{Replication-aligned convolution (RAConv)} goes even further by eliminating most of the RaS computation in addition.
RAConv performs a sequence of \{$Slide_f$, $MulFilter$, $Sum_ {f,c_i}$\} operations on $\pi_{RA}$-formatted inputs and returns $\pi_{CA}$-formatted results.
$Slide_f$ expands input ciphertexts with $f^2$ rotations and $MulFilter$ multiplies slided ciphertexts with plaintext filters using MulPt operations.
$Sum_{f,c_i}$ accumulates the intermediate results of $f^2\!\cdot\!c_i$ ciphertexts using only additions (AddCt) without costly rotation operations.
The resulting ciphertexts follow the $\pi_{CA}$ format, allowing the next CAConv to directly utilize them as input.

As the number of input ciphertexts ($n_i$) of RAConv has increased by the number of input channels $c_i$, additional rotation cost ($c_i\cdot(f^2 - 1)$ rotations) is required for $Slide_f$.
The increase in the $Slide_f$ cost undermines the performance enhancement from reduced rotations in RaS and IR.
We resolve this issue by reordering the operations in RAConv, which significantly reduces the $Slide_f$ cost.

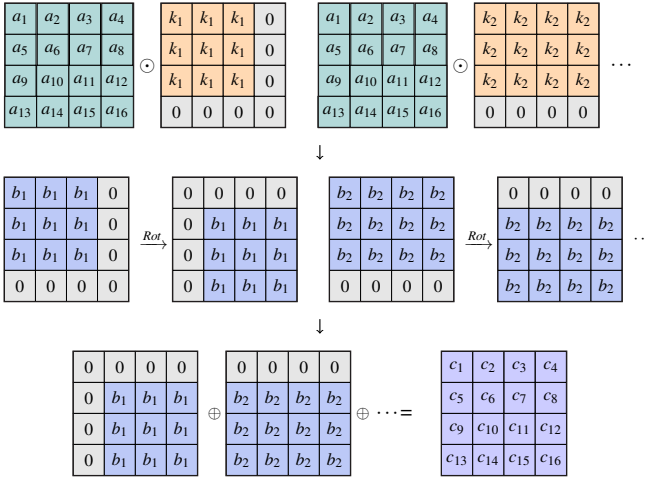
\begin{figure}[t]
        \centering
        \scalebox{0.55}{\begin{tikzpicture}[mymatrixenv]
    \matrix(m1)[mystyle2,font=\large]{
    $a_{1}$&$a_{2}$&$a_{3}$&$a_{4}$\\
    $a_{5}$&$a_{6}$&$a_{7}$&$a_{8}$\\
    $a_{9}$&$a_{10}$&$a_{11}$&$a_{12}$\\
    $a_{13}$&$a_{14}$&$a_{15}$&$a_{16}$\\
    };
    
    \node[right = -0.4em of m1, scale=1.2] (str1) {\large $\odot$};
    \draw[thick, black] (m1-1-1.north west) rectangle (m1-4-4.south east);

    \matrix(f1)[style3, right= -0.4em of str1,font=\large]{
    $k_{1}$&$k_{1}$&$k_{1}$&\myzero\\
    $k_{1}$&$k_{1}$&$k_{1}$&\myzero\\
    $k_{1}$ & $k_{1}$ & $k_{1}$&\myzero\\
    \myzero&\myzero&\myzero&\myzero\\
    };
    \draw[thick, black] (f1-1-1.north west) rectangle (f1-4-4.south east);

    \matrix(m2)[mystyle2,right= 1.5em of f1, font=\large]{
    $a_{1}$&$a_{2}$&$a_{3}$&$a_{4}$\\
    $a_{5}$&$a_{6}$&$a_{7}$&$a_{8}$\\
    $a_{9}$&$a_{10}$&$a_{11}$&$a_{12}$\\
    $a_{13}$&$a_{14}$&$a_{15}$&$a_{16}$\\
    };
    \node[right = -0.4em of m2, scale=1.2] (str2) {\large $\odot$};

    \draw[thick, black] (m2-1-1.north west) rectangle (m2-4-4.south east);

    \matrix(f2)[style3, right= -0.4em of str2,font=\large]{
    $k_{2}$&$k_{2}$&$k_{2}$&$k_{2}$\\
    $k_{2}$&$k_{2}$&$k_{2}$&$k_{2}$\\
    $k_{2}$&$k_{2}$&$k_{2}$&$k_{2}$\\
    \myzero&\myzero&\myzero&\myzero\\
    };
    \draw[thick, black] (f2-1-1.north west) rectangle (f2-4-4.south east);
    \node [right = 0em of f2, scale=1.2] (eq2) {\large $\cdots$};
\end{tikzpicture}}\\\vfill
        \scalebox{0.55}{\begin{tikzpicture}[mymatrixenv]
    \node(H) {};
    \node(L)  [below = 1em of H] {};
    \draw[->, thick] (H.south -| L.north) -- (L.north);
\end{tikzpicture}}\\\vfill
        \scalebox{0.55}{\begin{tikzpicture}[mymatrixenv]
    \matrix(m1)[mystyle2,font=\large, nodes={draw, fill=myblue2!30}]{
    $b_{1}$&$b_{1}$&$b_{1}$&\myzero\\
    $b_{1}$&$b_{1}$&$b_{1}$&\myzero\\
    $b_{1}$&$b_{1}$&$b_{1}$&\myzero\\
    \myzero&\myzero&\myzero&\myzero\\
    };
    
    \node[right = 0em of m1] (str1) {\large $\xrightarrow{Rot}$};

    \draw[thick, black] (m1-1-1.north west) rectangle (m1-4-4.south east);

    \matrix(f1)[mystyle1, right= -0.4em of str1,font=\large, nodes={draw, fill=myblue2!30}]{
    \myzero&\myzero&\myzero&\myzero\\
    \myzero&$b_{1}$&$b_{1}$&$b_{1}$\\
    \myzero&$b_{1}$&$b_{1}$&$b_{1}$\\
    \myzero&$b_{1}$&$b_{1}$& $b_{1}$\\
    };
    \draw[thick, black] (f1-1-1.north west) rectangle (f1-4-4.south east);

    \matrix(m2)[mystyle2,right= 1.5em of f1, font=\large,nodes={draw, fill=myblue2!30}]{
    $b_{2}$&$b_{2}$&$b_{2}$&$b_{2}$\\
    $b_{2}$&$b_{2}$&$b_{2}$&$b_{2}$\\
    $b_{2}$&$b_{2}$&$b_{2}$&$b_{2}$\\
    \myzero&\myzero&\myzero&\myzero\\
    };
    \node[right = 0em of m2] (str2) {\large $\xrightarrow{Rot}$};
    \draw[thick, black] (m2-1-1.north west) rectangle (m2-4-4.south east);

    \matrix(f2)[mystyle1, right= -0.4em of str2,font=\large,nodes={draw, fill=myblue2!30}]{
    \myzero&\myzero&\myzero&\myzero\\
    $b_{2}$&$b_{2}$&$b_{2}$&$b_{2}$\\
    $b_{2}$&$b_{2}$&$b_{2}$&$b_{2}$\\
    $b_{2}$&$b_{2}$&$b_{2}$&$b_{2}$\\
    };
    \draw[thick, black] (f2-1-1.north west) rectangle (f2-4-4.south east);


    \node [right = 0em of f2, scale=1.2] (eq3) {$\cdots$};
\end{tikzpicture}}\\\vfill
        \scalebox{0.55}{\begin{tikzpicture}[mymatrixenv]
    \node(H) {};
    \node(L)  [below = 1em of H] {};
    \draw[->, thick] (H.south -| L.north) -- (L.north);
\end{tikzpicture}}\\\vfill
        \scalebox{0.55}{\begin{tikzpicture}[mymatrixenv]

    \matrix(f1)[mystyle1, right= -0.4em of str1,font=\large,nodes={draw, fill=myblue2!30}]{
    \myzero&\myzero&\myzero&\myzero\\
    \myzero&$b_{1}$&$b_{1}$&$b_{1}$\\
    \myzero&$b_{1}$&$b_{1}$&$b_{1}$\\
    \myzero&$b_{1}$&$b_{1}$& $b_{1}$\\
    };
    \draw[thick, black] (f1-1-1.north west) rectangle (f1-4-4.south east);

    \node[right = 0em of f1] (str1) {\large $\oplus$};

    \matrix(f2)[mystyle1, right= -0.4em of str1,font=\large,nodes={draw, fill=myblue2!30}]{
    \myzero&\myzero&\myzero&\myzero\\
    $b_{2}$&$b_{2}$&$b_{2}$&$b_{2}$\\
    $b_{2}$&$b_{2}$&$b_{2}$&$b_{2}$\\
    $b_{2}$&$b_{2}$&$b_{2}$&$b_{2}$\\
    };
    \draw[thick, black] (f2-1-1.north west) rectangle (f2-4-4.south east);
    

    \node [right =-0.4em of f2, scale=1.2] (eq2) {$\oplus$};
    \node [right = -0.4em of eq2, scale=1.2] (eq3) {\large $\cdots$=};
    
    \matrix(ret)[style2, right=1.2em of eq3, font=\large]{
    $c_{1}$&$c_{2}$&$c_{3}$&$c_{4}$\\
    $c_{5}$&$c_{6}$&$c_{7}$&$c_{8}$\\
    $c_{9}$&$c_{10}$&$c_{11}$&$c_{12}$\\
    $c_{13}$&$c_{14}$&$c_{15}$&$c_{16}$\\
    };
    \draw[thick, black] (ret-1-1.north west) rectangle (ret-4-4.south east);
\end{tikzpicture}}
	\caption{Plaintexts inversely rotated for SISO reordering.}
	\label{fig:siso_lazy}
\end{figure}

\renewcommand{\arraystretch}{1.1}

\begin{algorithm}
\caption{\reviewertwo{Reordered SISO for RAConv}} \label{alg:lazy-SISO}
\textbf{Input} $ct_{i}$: input ciphertext, $W'$: filter plaintexts inversely rotated \\
\textbf{Output} $ct_{o}$: output ciphertext
\begin{algorithmic}[1] 
    \STATE{// \textbf{MulFilter}\&\textbf{Sum$_{c_{i}}$}($ct_i,W$)}
    \FOR {$j_1= 0 ,\ldots,f-1 $}
        \FOR {$j_2= 0 ,\ldots,f-1 $}
            \STATE $ct'[j_1,j_2] \mathrel{+}=$ MulPt($ct_i, W'[j_1,j_2]$)
        \ENDFOR
    \ENDFOR
    \STATE{// \textbf{Slide$_1$}\&\textbf{Sum$_f$}($ct'$)}
    \FOR {$j_1= 0 ,\ldots,f-1 $}
        \FOR {$j_2= 0 ,\ldots,f-1 $}
            \STATE $r \leftarrow w_i(j_1- \frac{f-1}{2}) + (j_2 - \frac{f -1}{2})$
            \STATE $ct_{o} \mathrel{+}=$ CRot($ct'[j_1,j_2], -r$)
        \ENDFOR
    \ENDFOR
    \STATE \textbf{return} $ct_{o}$
\end{algorithmic}
\end{algorithm}

\begin{algorithm}
\caption{\reviewertwo{CAConv and RAConv Fused Block}} \label{alg:CARAConv}
\textbf{Input} $ct_{i}$: input ciphertext vector, $W_{CA}, W_{RA}$: weight plaintexts for CAConv, RAConv, $n_{CA}$, $n_{RA}$: number of ciphertexts with data format $\pi_{CA}, \pi_{RA}$, $ct_k$ : temporary ciphertexet vectors\\
\textbf{Output} $ct_{o}$: output ciphertext vector
\begin{algorithmic}[1] 
    \FOR {$k= 0 ,\ldots, n_{CA}\text{-}1 $}
        \STATE $ct_1[k]$ $\leftarrow$ \textbf{Slide$_f$}($ct_{i}[k]$)
    \ENDFOR
    \FOR {$j= 0 ,\ldots, n_{RA}\text{-}1 $}
        \FOR {$k= 0 ,\ldots, n_{CA}\text{-}1 $}
            \STATE $ct_2[j, k] \mathrel{+}= $\textbf{MulFilter\&Sum$_f$}($ct_1[k],W_{CA}[j,k]$)
            \STATE $ct_3[j]$ $\leftarrow$ \textbf{RaS$_{c_{i}}$}($ct_2[j,k]$)
        \ENDFOR
        \STATE $ct_3[j]$ $\leftarrow$ \textbf{Square}($ct_3[j]$)
        \FOR {$l= 0 ,\ldots, n_{CA}\text{-}1 $}
            \STATE $ct_4[l] \mathrel{+}=$ \textbf{MulFilter}\&\textbf{Sum$_{c_{i}}$}($ct_3[j],W_{RA}[l,j]$)
        \ENDFOR
    \ENDFOR
    \FOR {$k= 0 ,\ldots,  n_{CA}\text{-}1 $}
    \STATE $ct_{o}[k] \leftarrow$ \textbf{Slide$_1$}\&\textbf{ Sum$_f$}($ct_4[k]$)
    \ENDFOR
    \STATE \textbf{return} $ct_{o}$
\end{algorithmic}
\end{algorithm}

\textbf{Reordered RAConv:}
We \emph{rearrange} the RAConv sequence
to $\{MulFilter, Sum_{c_i}, Slide_1, Sum_f\}$ under the observation that the order of sliding and filter multiplication can be reversed if we prepare filter plaintexts to be inversely rotated.
\reviewertwo{We describe the procedure of reordered SISO in Figure~\ref{fig:siso_lazy} and Algorithm~\ref{alg:lazy-SISO}.
Unlike $Slide_f$ that generates $f^2$ rotated ciphertext from a single input ciphertext, $Slide_1$ gathers $f^2$ ciphertexts into one by single rotations and addition.
}
By reordering these operations, we can perform rotations after $n_i$ input ciphertexts are accumulated as shown in Figure~\ref{fig:alternate2_nostride_lazy}, effectively reducing the number of rotations required for sliding from  $c_i\cdot(f^2 - 1)$ to $(f^2-1)$.

\begin{table*}[ht]
    \centering
\caption{\centering Cost of homomorphic convolutions. We compare our convolutions with \cite{lee_2022_low} when $n_{i}=n_{o}=1$ in their setting. ($c_n$: channel multiplexing in $C_a$ dimension), (m: channel multiplexing in $C_g$ dimension), (d: input repetition in R dimension).} 
\label{Rotation Complexity}
    \begin{tabular}{cccccccc}
        \toprule
        \textbf{Method}  & $n_{i}$ & $n_{o}$ & \textbf{Slide}  & \textbf{RaS} & \textbf{RaS$_g$} & \textbf{IR}   & \textbf{IR$_g$} \\ 
        \midrule
        \textbf{MP$\text{-}$Conv$_{lc}$~\cite{lee_2022_low}} &  1 & 1    & $f^2\text{-}1$   &  $\frac{mc_n}{d}\log({c_n})$ & $\frac{mc_n}{d}\log({m})$   & \multicolumn{2}{c}{$(mc_n\text{-}1\text{+}\log_2(d))$} \\
        \midrule
        \textbf{CAConv} & 1 & $\frac{mc_n}{d}$ & $ f^2\text{-}1$ 
        &  $\frac{mc_n}{d}\log({c_n})$ & $\frac{mc_n}{d}\log({m})$  &  0  & $\frac{mc_n}{d}\log({m})$\\
        \midrule
        \textbf{RAConv$_{Naive}$} & $\frac{mc_n}{d}$ & 1 & $\frac{mc_n}{d}(f^2\text{-}1)$& 0  &   $\log({m})$ & 0 & $\log({m})$\\
        \midrule
        \textbf{RAConv$_{Reorder}$} & $\frac{mc_n}{d}$ & 1 & $f^2\text{-}1$ & 0  &  $\log({m})$ & 0 & $\log({m})$\\
        \bottomrule
    \end{tabular}
\end{table*}

\reviewertwo{We also propose an inter-layer optimization aimed at reducing the memory footprint required for ciphertexts in CAConv and RAConv, as outlined in Algorithm~\ref{alg:CARAConv}. In homomorphic convolutions, the number of input, output, and temporary ciphertexts is determined by their data formats.
Our optimization utilizes aggressive forwarding which aims to avoid states where the intermediate data formats occupy large memory space (e.g. $ct_2, ct_3$ in Algorithm~\ref{alg:CARAConv}).
We denote the number of inputs for CAConv and RAConv as $n_{CA}$ and $n_{RA}$, respectively, to prevent confusion incurred by fusing these two convolutions. 
By fusing loops that yield $ct_2$ and $ct_3$ into $ct_4$, we forward each ciphertext from $ct_2, ct_3$ to subsequent operations until obtaining $ct_4$ (see line number 4-9). 
}
Our forwarding optimization retains the number of ciphertexts to $n_{CA}\cdot f^2$, which is much smaller than $n_{RA}\cdot f^2$.

\subsection{Data Formats for 2D Gap Packing} 
\label{subsec:2d_gap_packing}
\begin{figure}[t]
\hfil
        \subfloat[Convolution with Multiplexed packing]{\label{fig:hybrid_IAConv}
            \centering{
    	    \scalebox{0.6}{

    
   
    

\begin{tikzpicture}[mymatrixenv]
    \matrix(a1)[mystyle111,nodes={draw,minimum width=1cm}]{
    \gf$a^{(1)}$&\gggg$a^{(2)}$\\
    \ggg$a^{(3)}$&\ggggg$a^{(4)}$\\
    };
    \node[right = 0em of a1] (conv1) {$\xrightarrow{Conv_{lc}}$};

    \matrix(c1)[mystyle111, right = 0.2em of conv1,nodes={draw,minimum width=1.6cm}]{
    \mb$b^{(4k+1,1)}$&\mbb$b^{(4k+2,1)}$\\
    \mbbb$b^{(4k+3,1)}$&\mbbbb$b^{(4k+4,1)}$\\
    };
    
    \draw [->,thick] (c1-1-2.north) to [out=150,in=30] node [text width=2cm,midway,above]{RaS} (c1-1-1.north);
    \draw [->,thick] (c1-2-1.west) to [out=150,in=210] node [text width=5cm,midway,above]{} (c1-1-1.west);
   
    \node[right = 0em of c1] (ras1) {$\xrightarrow{}$};
    \matrix(c2)[mystyle111, right = 0.2em of ras1,nodes={draw,minimum width=1cm}]{
    \mb$c^{(1)}$ & \myhash\\
    \myhash   & \myhash\\
    };
    

    \node[right = 0em of c2] (bcast1) {$\xrightarrow{IR}$};
    \matrix(c3)[mystyle111, right = 0em of bcast1,nodes={draw,minimum width=1cm}]{
    \mb$c^{(1)}$ & \mbb$c^{(2)}$   \\
    \mbbb$c^{(3)}$      & \mbbbb$c^{(4)}$  \\
    };

\end{tikzpicture}}}}
      \hfil
      \par
      \hfil
        \subfloat[Convolution of 2D gap packing, data format (R, C) ]{\label{fig:hybrid_RAConv}
            \centering{
    	    \scalebox{0.6}{
    	        \begin{tikzpicture}[mymatrixenv]
    \matrix(a1)[mystyle111,nodes={draw,minimum width=1cm}]{
    \gf$a^{(1)}$&\gggg$a^{(2)}$\\
    \gf$a^{(1)}$&\gggg$a^{(2)}$\\
    };
    \node[right = 0em of a1] (conv1) {$\xrightarrow{CAConv}$};

    \matrix(c1)[mystyle111, right = 0.2em of conv1,nodes={draw,minimum width=1.6cm}]{
    \mb$b^{(2k+1,1)}$&\mbb$b^{(2k,1)}$\\
    \mbbb$b^{(2k+1,2)}$&\mbbbb$b^{(2k,2)}$\\
    };
    
    \draw [->,thick] (c1-1-2.north) to [out=150,in=30] node [text width=2cm,midway,above]{RaS} (c1-1-1.north);
   
    \node[right = 0em of c1] (ras1) {$\xrightarrow{ }$};
    \matrix(c2)[mystyle111, right = 0.2em of ras1,nodes={draw,minimum width=1cm}]{
    \mb$c^{(1)}$ & \myhash \\
    \mbbb$c^{(2)}$   & \myhash\\
    };
    
    \draw [<-,thick] (c2-1-2.north) to [out=120,in=60] node [text width=4cm,midway,above]{IR (mask and rotate)} (c2-1-1.north);

    \node[right = 0em of c2] (bcast1) {$\xrightarrow{}$};
    \matrix(c3)[mystyle111, right = 0em of bcast1,nodes={draw,minimum width=1cm}]{
    \mb$c^{(1)}$ & \mb$c^{(1)}$   \\
    \mbbb$c^{(2)}$      & \mbbb$c^{(2)}$  \\
    };
\end{tikzpicture}}}}
        \hfil
      \par
        \hfil
        \subfloat[Convolution of 2D gap packing, data format (C, R) ]{\label{fig:hybrid_RAConv2}
            \centering{
    	    \scalebox{0.6}{
    	        \begin{tikzpicture}[mymatrixenv]
    \matrix(a1)[mystyle111,nodes={draw,minimum width=1cm}]{
    \gf$a^{(1)}$&\gf$a^{(1)}$\\
    \gggg$a^{(2)}$&\gggg$a^{(2)}$\\
    };
    \node[right = 0em of a1] (conv1) {$\xrightarrow{RAConv}$};

    \matrix(c1)[mystyle111, right = 0.2em of conv1,nodes={draw,minimum width=1.6cm}]{
    \mb$b^{(2k+1,1)}$&\mbb$b^{(2k+1,2)}$\\
    \mbbb$b^{(2k,1)}$&\mbbbb$b^{(2k,2)}$\\
    };
    \tikzstyle{line} = [draw, -latex']
    \draw [->,thick](c1-2-2.east) to [bend right=30] node[pos=0.2,text width=1cm,midway,right]{RaS} (c1-1-2.east);
   
    \node[right = 2em of c1] (ras1) {$\xrightarrow{}$};
    \matrix(c2)[mystyle111, right = 0.2em of ras1,nodes={draw,minimum width=1cm}]{
    \mb$c^{(1)}$ & \mbbb$c^{(2)}$ \\
    \myhash  & \myhash\\
    };
    
    \draw [<-,thick] (c2-2-2.east) to [bend right=30] node [text width=1cm,midway,right]{IR} (c2-1-2.east);

    \node[right = 1.2em of c2] (bcast1) {$\xrightarrow{}$};
    \matrix(c3)[mystyle111, right = 0em of bcast1,nodes={draw,minimum width=1cm}]{
    \mb$c^{(1)}$ & \mbbb$c^{(2)}$   \\
    \mb$c^{(1)}$      & \mbbb$c^{(2)}$  \\
    };
\end{tikzpicture}}}}
      \hfil
	\caption{The procedure of convolution on single pixel of image with gap size $g\text{=}2$. $a_i^{(j)}$ denotes the $i$-th element in the $j$-th channel of an input image tuple $a$. $b^{(mk+n,l)}$ represents the channels are accumulated by stride m $\sum_{k=1}^{\frac{c_i}{m}}b^{(mk+n,l)}$ \label{fig:hybridconv}}\vspace{-0.05cm}
\end{figure}
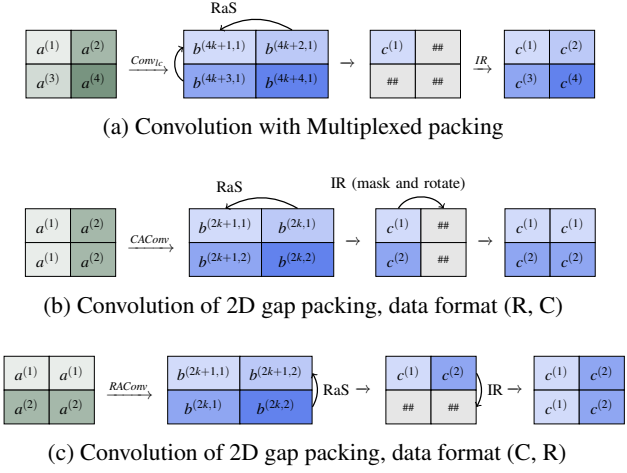

We propose a gap packing method that collaboratively employs duplication and channel multiplexing to alleviate the high cost of the repacking process.
In the innermost two dimensions, referred to as \emph{2D gap packing}, $C_g$ and $R_g$ of $\pi_{CA}$ and $\pi_{RA}$ represent duplication and channel multiplexing, respectively.
Unlike $MP\text{-}Conv_{lc}$, which maintains a single $C_g$ for convolution, $\pi_{CA}$ and $\pi_{RA}$ act together in a complementary manner such that each convolution transforms $C_g$ ($R_g$) into $R_g$ ($C_g$).
Our key observation is that introducing this heterogeneity can significantly reduce the number of rotations being invoked.
Specifically, in $MP\text{-}Conv_{lc}$, image repacking inside the gap (IR$_g$) spends O($g^2$) rotations to maintain the gap packing by $C_g$, when $g$ denotes the gap width (height).
In contrast, when using our 2D gap packing, IR only requires O($\log g^2$) rotations in the $R_g$ direction as shown in Figure~\ref{fig:hybrid_RAConv} and Figure~\ref{fig:hybrid_RAConv2}.

\subsection{Complexity Analysis}
We describe the complexity of rotation counts of convolution algorithms in Table~\ref{Rotation Complexity}.
We first start with the baseline~\cite{lee_2022_low}, and  denote $c_n$, $m$ and $d$ as $|C_a|$, $|C_g|$ and $|R|$ of $MP\text{-}Conv_{lc}$.
In CAConv, d turns into $|R_g|$ as we relocate input repetition to the gap dimension, then RAConv proceeds on the output of CAConv.
Except for the RaS of CAConv, RaS and IR does not require any rotation for our convolutions.
As elaborated in Section~\ref{sec:Convolution Algorithms}, our convolution does not require any rotations except for RaS of CAConv aside from gap.
By reorganizing the order of sliding rotation, Slide of RAConv$_{Reorder}$ requires $\frac{mc_n}{d} \times$ fewer rotations compared to RAConv$_{Naive}$
We also separately present rotation amount to retain our 2D gap packing method ($RaS_g$ and $IR_g$).
Overall, our construction halves the rotation cost of RaS and avoids high rotation cost of IR.

We highlight that $n_{i}$ and $n_{o}$ has direct impact on the total number of activation function and bootstrapping.
For instance, $n_i$ of RAConv is typically larger than $n_i$ of CAConv, increasing the number of activation function or bootstrapping on $\pi_{RA}$.
Unlike square activation function which has minor impacts on overall performance, HCNN performance is critically dependent on the number of bootstrapping. Therefore, we conduct bootstrapping when the number of ciphertext is minimal, particularly in the format of $\pi_{CA}$. Furthermore, during end-to-end network implementation, we fine-tune $(m, d)$ on each block, aiming for the least total execution time for rotation and bootstrapping. An in-depth performance analysis of the choice of $(m, d)$ is provided in Section~\ref{subsec:parameter}.

\begin{figure}[t]
        \subfloat[CAConv with PRCR]{\label{fig:Slicing_CAConv}
            \centering
    	        \includegraphics[width=0.98\columnwidth]{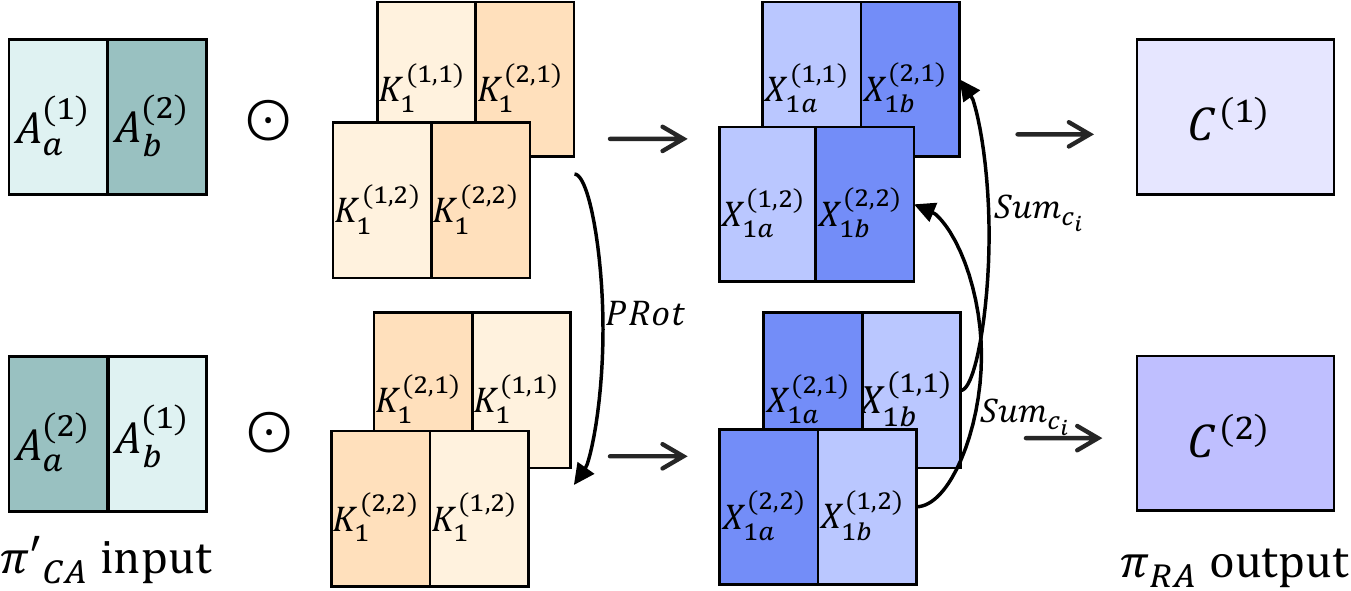}}
             \\
        \subfloat[RAConv with PRCR]{\label{fig:Slicing_RAConv}
            \centering
    	        \includegraphics[width=0.98\columnwidth]{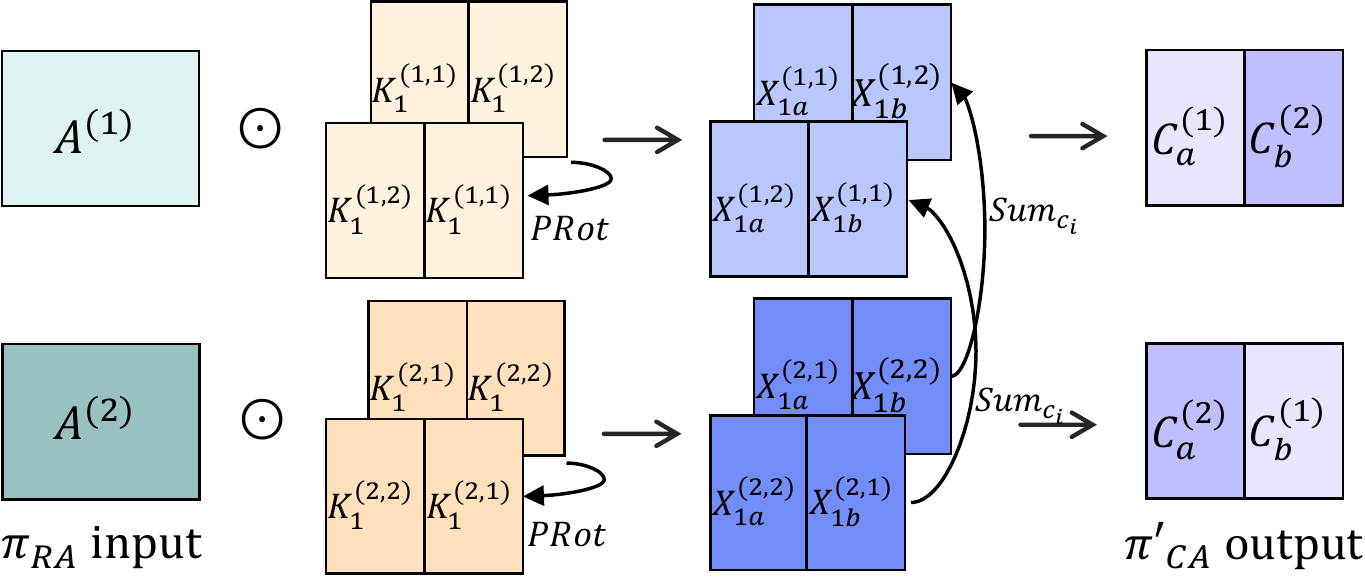}}
    	        
	\caption{Simplified convolution processes with PRCR. We assume $c_i=c_o=2$ but with images per ciphertext $c_n$ and filter size $f$ equals one for this figure. Ciphertexts are filled with subscript \emph{a} and \emph{b} represents upper half and lower half of an image, respectively
    (e.g., $\pi'_CA$ of (a) has single image, but each composed of different half of channel image). PRCR reuses weight plaintexts through plaintext rotation (PRot).  \label{fig:ImageSlicing}}
\end{figure}

\subsection{Plaintext size reduction through Image to ciphertext rearrangement}
\label{subsec:Image Slicing}

When extending HCNN to high-resolution images (e.g., 224$\times$224 images in ImageNet), HCNN encounters a significant surge in memory footprint.
This spike is especially pronounced in the weight plaintext, with each filter element occupying the fragment sized $w_ih_i$ slots as shown in Figure~\ref{fig:siso_stride1}.
In total, filters require $w_ih_if^2c_ic_o$ slots to produce weight plaintexts.
For instance, when running ResNet-18 on ImageNet, weight plaintexts alone occupy a substantial 364.8GB of memory space (see Table~\ref{tb:footprint}). This exceeds the memory capacity of a single cutting-edge GPU, limited by the current memory technology.

We propose an optimization named PRCR (\textbf{P}laintext size \textbf{R}eduction through the image to \textbf{C}iphertext \textbf{R}earrangement) to mitigate the data expansion of plaintexts.
Our method involves splitting an image into multiple image segments along the row direction.
Thus, the data format of CAConv is adjusted to $\pi_{CA'} = \{C_a,S,H',W,R_g,C_g\}$, where $S$ denotes the dimension of sub-images and $H'$ denotes the dimension of rows of a image segment.
This rearrangement can reduce the fragment size of a filter element to the image segment size.
Furthermore, These fragments are organized circularly, allowing the pieces to be reused by permutation.
In Figure~\ref{fig:Slicing_CAConv}, we depict PRCR method when $slot\text{=}w_ih_i$, $f\text{=}1$, and $c_i\text{=}c_o\text{=}2$.
CAConv reuses a single weight plaintext $|S|$ times, by multiplying with an input ciphertext and rotating by a fragment size.
Subsequent Sum$_{c_{i}}$ accumulates input channels, restoring image-size fragments and generating replication-aligned ciphertexts.
While $\pi_{RA}$ remains unchanged, PRCR rearranges weight plaintexts into smaller fragments, placing different output channels circularly.
As CAConv, RAConv reuses the weight plaintexts $|S|$ times with multiplication and rotation and returns $\pi_{CA'}$.

PRCR can be independently applied along with any form of packing techniques to reduce plaintext memory.
The extra computation of PRot is not significant to the overall performance, 
because PRot is much cheaper than other FHE operations as presented in Table~\ref{tb:Benchmark}.
For our implementation, we use $|S| = 8$ for ImageNet. However, as GPU memory is capable of CIFAR-10 workloads, we do not apply PRCR to CIFAR-10.

\subsection{ResNet Construction with HyPHEN}
\label{subsubsec:Architecture}
\begin{figure}[t]
\centering
\includegraphics[width=0.95\columnwidth]{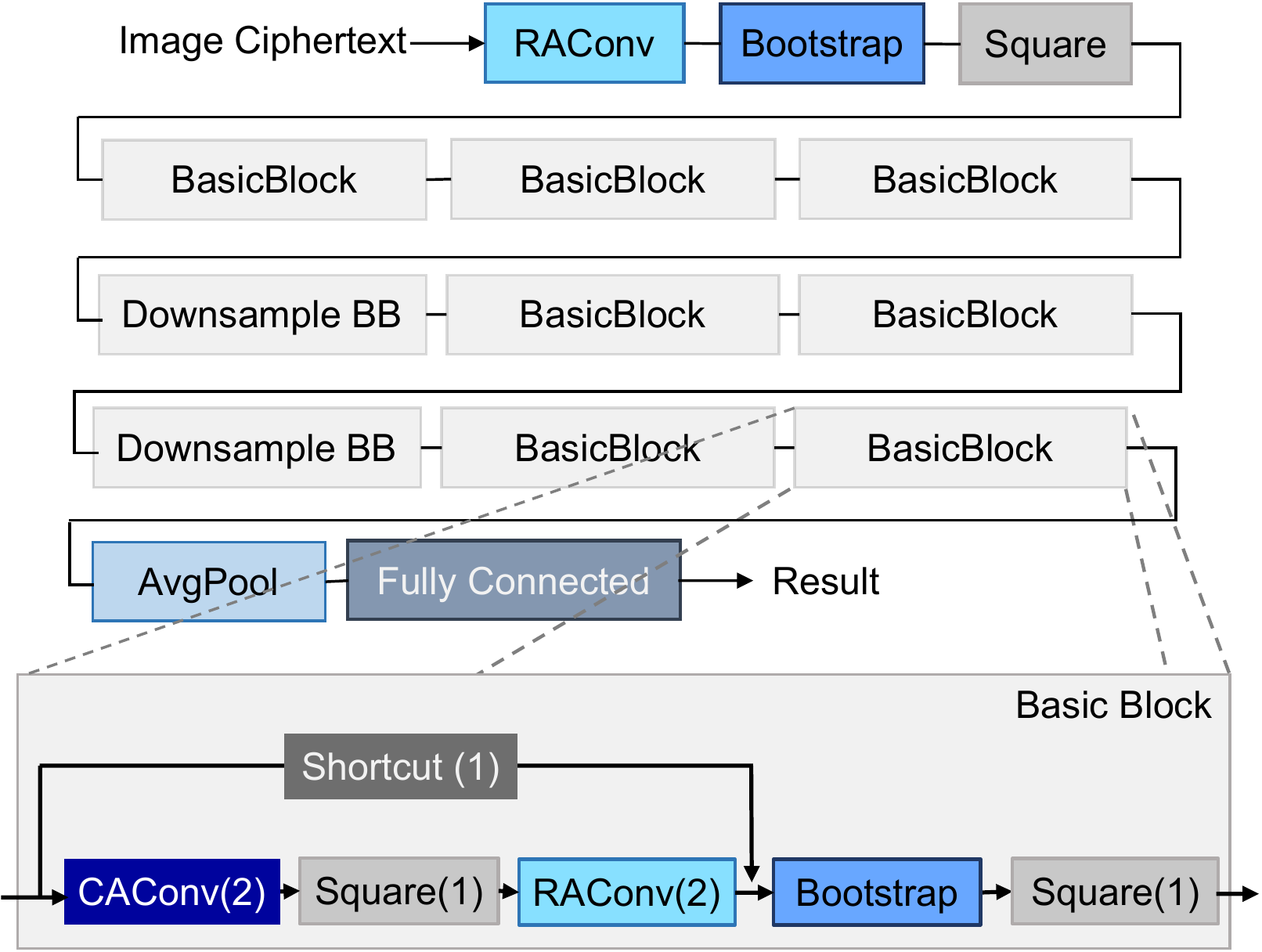} 
\caption{ResNet basic block built on HyPHEN. The level consumption per block is written in each parentheses. In the downsampling block, pointwise convolution is added to the critical path. Otherwise, a simple shortcut is added. \label{HyPHEN_architecture}}
\end{figure}

HyPHEN combines RAConv, 2D gap packing and PRCR to build the entire CNN model.
Figure~\ref{HyPHEN_architecture} illustrates the ResNet implemented in HyPHEN.
There are three considerations when deciding the placement of operations.
First, it is effective to place bootstrapping after RAConv, and not CAConv, due to the smaller number of ciphertexts involved.
Second, to match the levels between the shortcut path and the main CAConv-RAConv path, bootstrapping should be positioned either before the divergence of residual connections or after their convergence.
Lastly, it is beneficial to perform convolutional layers at the lowest level possible because the complexity of FHE operations such as rotation, is proportional to the ciphertext level $l$.

Putting everything together, our ResNet basic block consumes a total of 6 levels.
The level consumption of each layer is represented in the parentheses of each block.
CAConv and RAConv use 2D gap packing and consume one level for each of SISO and IR.
For convolution with $m>1$ need level consumption for multiplication with mask. 
We adopted AESPA for activation which consumes one level.
AESPA is a quadratic polynomial with different coefficients for each channel.
During inference, we fuse the coefficients into nearby layers, resulting in a simple square function $x^2$ for activation.
We set the ciphertext level after bootstrapping ($L'$ in Section~\ref{subsec:FHE}) to six and perform bootstrapping every time when the level becomes zero.
\section{Evaluation}
\label{sec:Evaluation}

\subsection{Experimental Setup}
\label{subsec:Benchmarks}
We conducted HCNN inference in both CPU and GPU environments using the RNS-CKKS library, HEaaN~\cite{heaan_library}.
The CPU system is equipped with two AMD EPYC 7452 CPUs running at 2.35GHz (32 cores per socket) and 480GB of DRAM.
GPU experiments were carried out on the same system with an additional NVIDIA A100 GPU with 80GB of memory.
In our HCNN inference experiments, we used ResNet-20/32/44 for CIFAR-10~\cite{krizhevsky_2009_cifar} and ResNet-18 for ImageNet~\cite{russakovsky_2015_imagenet} datasets.
Training was conducted with AESPA on PyTorch under normal supervised setting following the original paper.
We utilized the Kaiming-normal initialization method to initialize convolution and fully connected layers. After the training process, we applied a fusion technique to combine the weights and biases of batch normalization (BN) and the coefficients of AESPA with convolution layers.
Our RNS-CKKS parameters satisfy a 128-bit security level~\cite{cheon_2019_hybrid} with a polynomial degree $N\!=\!2^{16}$ and a hamming weight of 192.
Table~\ref{tb:resnet 20 parameter} displays the parameters used in the convolution layers of ResNet-20/32/44/18. All the parameters ($c_i, c_o, w_i, w_o, f, s)$ are determined following the original ResNet paper~\cite{he_2016_deep}. 
Execution time measurement begins once all weight plaintexts and input ciphertexts are loaded into either host or device memory. It ends once the encrypted inference result is returned.
The client's decrypt and encrypt processes are excluded as they are not considered critical bottlenecks in HCNN inference.

\setlength{\tabcolsep}{2.2pt}
\begin{table*}[t]
\centering
\caption{Model architecture detail of ResNet-20/32/44 for CIFAR-10 and ResNet-18 for ImageNet. Downsampling Convolution with stride 2 is represented as dsconv. Pointwise convolution with filter size 1 is denoted as pconv.\label{tb:resnet 20 parameter} }
{\small
\begin{tabular}{l|c|ccc|ccc|c|ccc|ccc|ccc}
\toprule[1.0pt]
  & \multicolumn{7}{c|}{\textbf{ResNet-20/32/44}} & \multicolumn{10}{c}{\textbf{ResNet-18}}\\
  & Layer1 & \multicolumn{3}{c|}{Layer2} & \multicolumn{3}{c|}{Layer3} & Layer1 & \multicolumn{3}{c|}{Layer2} & \multicolumn{3}{c|}{Layer3} & \multicolumn{3}{c}{Layer4}\\  
  & \textbf{conv} &  \textbf{dsconv} & \textbf{pconv} &\textbf{conv} &  \textbf{dsconv} & \textbf{pconv} &\textbf{conv} & \textbf{conv} &  \textbf{dsconv} & \textbf{pconv} &\textbf{conv} &  \textbf{dsconv} & \textbf{pconv} &\textbf{conv} & \textbf{dsconv} & \textbf{pconv} &\textbf{conv} \\  
\midrule[0.6pt]
$c_i$ &16&16&16&32&32&32&64&64&64&64&128&128&128&256&256&256&512\\
$c_o$ &16&32&32&32&64&64&64&64&128&128&128&256&256&256&512&512&512\\
$w_i$ &32&32&32&16&16&16&8&56&56&56&28&28&28&14&14&14&7\\
$w_o$ &32&16&16&16&8&8&8&56&28&28&28&14&14&14&7&7&7\\
$f$ &3&3&1&3&3&1&3&3&3&1&3&3&1&3&3&1&3\\
$s$ &1&2&2&1&2&2&1&1&2&2&1&2&2&1&2&2&1\\
\bottomrule[1.0pt]
\end{tabular}
}
\end{table*}
\setlength{\tabcolsep}{6pt}

\begin{figure}[t]
\centering
\includegraphics[width=0.98\columnwidth]{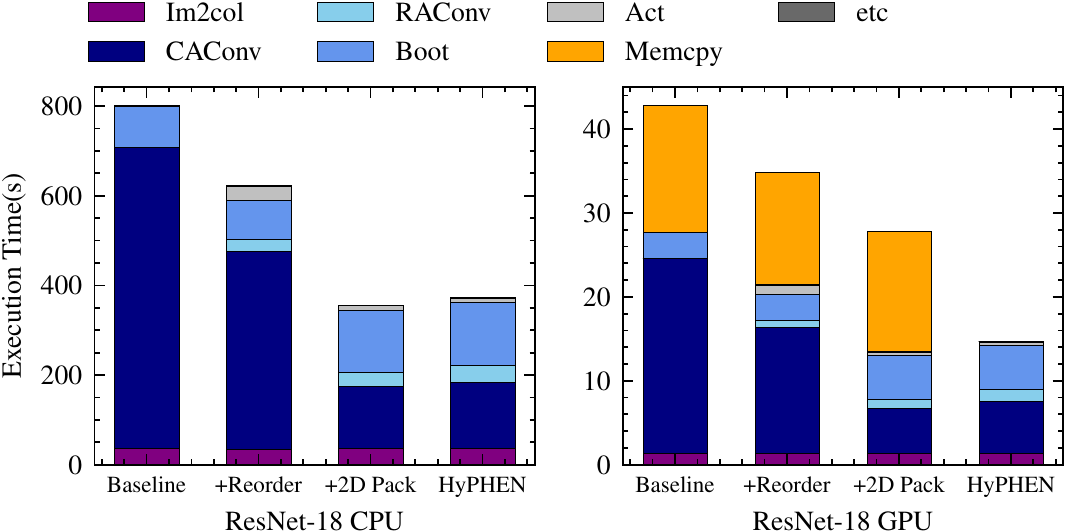} 
\caption{HCNN execution time of ResNet-18 (ImageNet) on our CPU and GPU systems.
\label{sensitivity study}}
\end{figure}

\renewcommand{\arraystretch}{1.0}
\begin{table}[t]
\caption{Memory footprint of ResNet-18 on ImageNet. \cite{lee_2022_low} is an extended implementation to ImageNet. Baseline is \cite{lee_2022_low} augmented with AESPA, where HyPHEN improves it with PRCR.\label{tb:footprint}}
\centering
{\small
\begin{tabular}{ccc}
\toprule
\textbf{Implementation} & \textbf{Filter plaintexts} & \textbf{Evaluation keys} \\ 
\midrule
    \cite{lee_2022_low} & 285.1GB     & 69.7GB        \\
    Baseline & 364.8GB     & 11.1GB      \\
    \textbf{HyPHEN} & {\color{white} 0}45.6GB     & 11.1GB      \\
        \bottomrule
\end{tabular}
}
\end{table}

\setlength{\tabcolsep}{4pt}
\renewcommand{\arraystretch}{1}
\begin{table*}[t]
\centering
    \caption{HyPHEN inference time of a single CIFAR-10 image using ResNet20/32/44 and a single ImageNet image using ResNet-18 on CPU and GPU. As FC and pooling layers have a tiny execution time, we gather them at Others. For ResNet-18, we implement the initial downsampling convolution by im2col, which is also summed in Others.\label{tb:execution time}}
    {\small
    \begin{tabular}{lcccc|cccc}
    \toprule[1.0pt]
    \multirow{2}{*}
    {\begin{tabular}{@{}c@{}} 
    \textbf{Execution}\\ \textbf{time} (s)\end{tabular}} & \multicolumn{4}{c|}{\textbf{CPU} (64 threads)}    & \multicolumn{4}{c}{\textbf{GPU}}                        \\
     &  ResNet-20 & ResNet-32 & ResNet-44 & ResNet-18\textsuperscript{1} & ResNet-20 & ResNet-32 & ResNet-44 & ResNet-18\textsuperscript{2} \\
    \midrule[0.4pt]
    \textbf{CAConv}   & 12.85    &   19.39   &  24.42    &  148.33   & {\color{white} 0}0.46            &       {\color{white} 0}0.66       &       0.87      & 6.22 \\
    \textbf{RAConv}   & {\color{white} 0}1.68     &   {\color{white} 0}3.02    &   {\color{white} 0}4.01    &  {\color{white} 0}37.16   & {\color{white} 0}0.06            &       {\color{white} 0}0.09       &       0.13      &  1.37 \\
    \textbf{Bootstrap}   &  21.40    &   34.23   &  45.90    &  140.57    & {\color{white} 0}0.83            &     {\color{white} 0}1.32        &      1.83      & 5.32 \\
    \textbf{Activation}   &  {\color{white} 0}1.41     &  {\color{white} 0}2.41     &  {\color{white} 0}3.27     &  {\color{white} 00}9.65    & {\color{white} 0}0.05            &     {\color{white} 0}0.09        &      0.12      & 0.36 \\
    \textbf{Others}           &  {\color{white} 0}0.23     &  {\color{white} 0}0.30     &  {\color{white} 0}0.44     &  {\color{white} 0}37.70     &\textless 0.01   & \textless 0.01 & 0.01 & 1.42 \\
    \midrule[0.4pt]
    \textbf{Total}  &  37.57 $\pm$ 0.7   & 59.35 $\pm$ 0.9   & 78.04 $\pm$  0.8  & 373.41 $\pm$ 3.9    & 1.40 $\pm$ 0.04 & 2.17 $\pm$ 0.04  & 2.96 $\pm$ 0.04 & 14.69 $\pm$  0.01 \\
    \bottomrule[1.0pt]
    \multicolumn{9}{l}{\small \textsuperscript{1, 2} ResNet-18 experiments are conducted on the ImageNet dataset.}
    \end{tabular}
    }
\end{table*}
\setlength{\tabcolsep}{6pt}

\setlength{\tabcolsep}{3pt}
\renewcommand{\arraystretch}{1.0}
\begin{table}[t]
\caption{Comparison of the unencrypted classification accuracy (Backbone) and HCNN classification accuracy (HyPHEN) of the ResNet models. While ResNet-20/32/44 results are accuracy of CIFAR-10, ResNet-18 is the accuracy of ImageNet classification.   \label{tb:accuracy}}
\centering
{\small
\begin{tabular}{ccccc}
\toprule[1.0pt]
\mc{\textbf{Top-1 Acc (\%) }}         & \mc{\textbf{ResNet-20}} & \mc{\textbf{ResNet-32}} & \mc{\textbf{ResNet-44}} & \mc{\textbf{ResNet-18}}  \\ 
\midrule[0.6pt]
    \textbf{Backbone} & 92.18     & 93.36     & 94.04  & 65.25     \\
    \textbf{HyPHEN} & 92.17     & 93.35     & 94.08  & 65.25   \\
        \bottomrule[1.0pt]
\end{tabular}
}
\end{table}
\setlength{\tabcolsep}{6pt}

\subsection{Impact of HyPHEN's components on performance}
\label{subsec:Sensitivity}
To analyze the effectiveness of each component of HyPHEN, we gradually apply alternation between CAConv and RAConv with reordering (+ Reorder), 2D gap packing (+ 2P), and PRCR (HyPHEN). 
We tested with the CPU/GPU implementations of ResNet-18 for ImageNet.
The results are shown in Figure~\ref{sensitivity study}.
By Alternating between CAConv and RAConv with Reorder, we achieved a 1.29$\times$ speedup, and an additional 1.75$\times$ speedup due to 2D gap packing in CPU.
The benefits of both Reorder and 2D gap packing also extend to GPU, resulting a 2.05$\times$ speedup of computation time, excluding memory access latency.
Meanwhile, PRCR results in 1.89$\times$ latency reduction in GPU, but leads to a slight slowdown (4.6\%) in CPU due to the added computation of plaintext permutation.
Although PRCR reduces the aggregate size of weight plaintexts by $s_i = 8$ times, it does not lead to latency reduction for our CPU system since its memory capacity can accommodate the whole working set (see Table~\ref{tb:footprint}). 
However, for the GPU system, PRCR enables the entire working set to fit in the 80GB GPU memory, reducing latency by eliminating the runtime overhead of copying data from the host CPU to the GPU (Memcpy).
In summary, PRCR offers benefits of substantial memory footprint reduction and latency reduction, especially for memory-constrained cases.

\subsection{Execution Time Breakdown}
\label{subsec:Executiontime}
Table~\ref{tb:execution time} displays the runtimes of ResNet-20/32/44 for the inference of a single CIFAR-10 image and ResNet-18 for a single ImageNet image.
Our ResNet-20/32/44 implementations on GPU take a few seconds to complete.
While the majority of inference time for ResNet20/32/44 is spent on bootstrapping, for ResNet-18, 49.6\% (CPU) and 51.7\% (GPU) of inference time is spent on convolution. This is because ResNet-18 has four times more channels than ResNet-20/32/44.
Table~\ref{tb:execution time} also demonstrates that our RAConv, which replaces about half of the convolution layers effectively reduces the overall runtime of the convolutional layers.

\renewcommand{\arraystretch}{1}
\begin{table*}[ht]
\centering
\caption{Runtime for the ResNet instances with different $(m,d)$ parameters and packing strategies. $^\dag$ We decompose rotation indices of unloaded evks  into loaded evks. Effective rotations, the number of rotations after decomposition are represented as eff.total. \label{tb:Rot and Boot}}
\Scale[0.75]{
\begin{tabular}{cc|cccc|cccc|c|c|c*{1}{d{3.1}}}
\toprule[1.0pt]
 \multirow{2}{*}{Model} &\multirow{2}{*}{Packing} & \multicolumn{4}{c|}{\textbf{(m,d)}} & \multicolumn{5}{c|}{\textbf{Conv Rotations}} & \multirow{2}{*}{\textbf{Boot}} & \multirow{2}{*}{\textbf{CPU Runtime}(s)} \\
                          & &   L1    &     L2                &     L3                &     L4               & SISO  & RaS & IR  & total & eff. total$^\dag$  &                       &                          \\ 
\midrule[0.4pt]
\multirow{3}{*}{ResNet-20} 
& Baseline&                    &                   &                    &           & 152   & 924   & 800   & 1876 & 3638  & 10 &   68.58 $\pm$ 1.0     \\
&Optimal    & (1,2)              & (2,4)             & (4,8)              &   -        & 152   & 580   & 187    & 919 & 1002   & 10     & \bf 37.57 $\pm$ 0.7     \\
&Min Rot    & (1,2)              & (1,8)             & (2,16)             &   -        & 240   & 407   & 142    & 789 & 881   & 15     & 44.51 $\pm$ 0.5  \\
\midrule[0.4pt]
\multirow{3}{*}{ResNet-18} & Baseline &                    &                   &                    &                 & 536   & 32384   & 4669    & 37589 & 43672  & 38     & 802.08 $\pm$ 7.6    \\ 
&Min Boot  & (1,1)              & (4,1)             & (16,1)              & (64,1)          & 536   & 17920  & 9544     & 28000 & 30072  & 38     &   623.11 $\pm$ 12.3\\
&Optimal   & (1,1)              & (2,2)             & (4,4)              & (8,8)          & 1024   & 4512   & 1823     & 7359 & 9095   & 65     &  \bf 356.97 $\pm$ 5.9     \\
\bottomrule[1.0pt]                    
\end{tabular}
}
\end{table*}

\subsection{Accuracy}
\label{subsec:Accuracy}
In Table~\ref{tb:accuracy}, we measured the classification accuracies of the validation set for CIFAR-10 running ResNet models with HyPHEN.
Near-zero accuracy degradation ($\leq$ 0.01\%) is observed for ResNet-20/32/44.
HyPHEN proves to be more robust to accuracy degradation than \cite{lee_2022_low}, which exhibits 0.09\% to 0.21\% accuracy degradation for ResNet-20/32/44 on CIFAR-10. 
The difference in accuracy drop can be explained by whether the original network is executed as is (using AESPA) or an approximation has been made (using ReLU approximation).
HyPHEN shows the same accuracy as the backbone accuracy even for a wider network (ResNet-18) and a bigger dataset (ImageNet).

\subsection{Parameter Study}
\label{subsec:parameter}
We conducted a parameter study to determine the optimal 2D gap packing setting for ResNet models to minimize latency. The choice of $(m, d)$ is crucial as it determines the number of rotations and bootstrappings to run the networks and thus execution time. Table~\ref{tb:Rot and Boot} shows the representative (m, d) instances, along with the resulting operation counts and execution times. We only present the $(m, d)$ of CAConv for simplicity, as $m$ and $d$ are exchanged at RAConv, and we omit ResNet-32/44 as ResNet-20/32/44 share the same optimal points.

In ResNet-20, \cite{lee_2022_low} uses input repetition because the size of the input tensor in the first layer (32 $\times$ 32 $\times$ 16) is smaller than the \ctxt slots ($2^{15}$). 
To avoid input repetition, we start with $(m, d)= (1, 2)$ in ResNet-20, instead of using input repetition. In ResNet-18, we start with the default $(m, d) = (1, 1)$. When input repetition occurs, it is more efficient to increase $d$ instead, but further increasing $m$ or $d$ does not yield better performance because it leads to more bootstrapping, as shown in our proposed architecture (see Figure~\ref{HyPHEN_architecture}). As the input ciphertexts go through the downsampling layer, $m\cdot d$ gets quadrupled and the size of the intermediate tensor gets halved. 2D gap packing, which doubles $m$ and $d$ every downsampling layer demonstrates optimal performance, corresponding to the minimal bootstrapping setting.
In ResNet-18, the minimal bootstrapping setting shows better performance than MP-CAConv. However, thanks to the flexibility of 2D gap packing we can find more efficient settings. The optimal setting requires 27 more bootstrappings and 20641 fewer rotations than the one with the minimum bootstrapping. 2D gap packing method helps balance the amount of rotation and bootstrapping to derive better performance.

\begin{figure}[t]
\centering
\includegraphics[width=0.98\columnwidth]{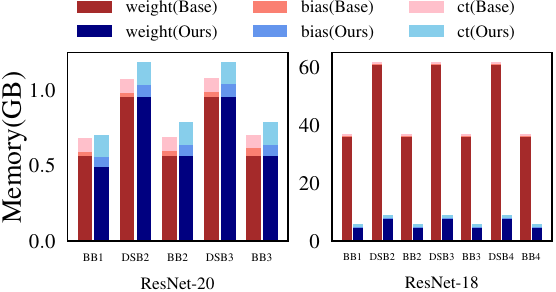} 
\caption{
Total memory size (GB) of each object in residual blocks.
We abbreviated Down-sampling Block and Basic Block to DSB and BB with each layer number as a postfix number.
Weight plaintexts, bias plaintexts, and input ciphertexts configure the memory footprint of each residual block.
Base and Ours represent the baseline and HyPHEN, respectively.
\label{memory study}}
\end{figure}

\begin{table}[ht]
\centering
\caption{FHE parameter settings. dnum is tuned to support 16, 6 levels required in \texttt{Set$_{lc}$}, and \texttt{Set$_{hyp}$}. Each ciphertext and plaintext memory size is represented when bootstrap refresh the level. \label{tb:parameter set} }
\Scale[0.7]{
\begin{tabular}{c c c c c c c}
\toprule[1.0pt]
  & $\mathbf{L+1}$ & $\mathbf{L'}$ & \textbf{Dnum} & \textbf{Ctxt (MB)} & \textbf{Ptxt (MB)} & \textbf{Evk (MB)}\\  
\midrule[0.6pt]
    \texttt{Set$_{lc}$} & 32 & 16     & 32    & 17     & 8.5 & 1056 \\
    \texttt{Set$_{hyp}$} & 24 & 6     & 6     & 10     & 5 & 168  \\
        \bottomrule[1.0pt]
\end{tabular}
}
\end{table}

\subsection{Training Details}
\label{sec:appendix_training}                                                                                           
The Models used in this paper are all trained using PyTorch~\cite{paszke_2019_pytorch}.
For ResNet-18 and 20, our training settings are mostly identical to AESPA; specifically, networks are trained for 200 epochs using an SGD optimizer.
We also use soft labels as in \cite{park_2022_aespa} to achieve higher accuracy. 
For ResNet-32 and 44, we use knowledge distillation~\cite{hinton_2015_knowledge} to enhance the accuracy, using pre-trained ResNet-32/44 with 93.4\% and 94.1\% accuracies as teacher models.
We adopt additional $l_2$ loss ($L_{kd} = \|f_t-f_s\|_2^2$) for distillation and trained for 240 epochs using the SGD optimizer.

\subsection{Memory Footprint Analysis}
\label{app:Memory requirement}

Memory-capacity requirement for HCNN depends on FHE parameters and data representations, such as packing schemes.
In FHE, data size expands during encoding and encryption procedures.
The resulting plaintexts and ciphertexts are typically orders of magnitude larger than the original messages.
Table~\ref{tb:parameter set} presents the actual size of ciphertexts, plaintexts, and evaluation keys on two FHE parameter settings.
We denote the RNS-decomposition number introduced in \cite{hen_2020_betterbt} as Dnum.
Given N, the degree of a cyclotomic polynomial ring, a large dnum increases L, the maximum level of a ciphertext.
We assume \texttt{Set$_{lc}$} is a setting used in \cite{lee_2022_low}.
As \cite{lee_2022_low} approximates ReLU with a high degree polynomial for activation, \texttt{Set$_{lc}$} adopts the maximum dnum to have $L'=16$.
\texttt{Set$_{hyp}$} is the parameter set used in this work.
Adopting AESPA allows us to select a smaller RNS-CKKS parameter ($L'=6$), as activation consumes one level. 
Certain FHE operations, such as MulCt, Rotate, and Conjugate, require the key-switching procedure.
Evk denotes the public evaluation key used during this process.
The size of a single Evk is 1,056MB and 176MB in \texttt{Set$_{lc}$} and \texttt{Set$_{hyp}$}, respectively.
For bootstrapping, one relinearization key for MulCt, one conjugation key, and 48 rotation keys are required.
Frequently used rotation keys for Slide are loaded for convolution. 
For instance, loading 66 unique Evks in ResNet-18 takes up 69.7GB and 11.1GB in \texttt{Set$_{lc}$} and \texttt{Set$_{hyp}$}, respectively.
Other irregular rotation keys used in IR are not loaded;
instead, these rotation indices are synthesized using the already loaded key indices.

Once FHE parameters are determined, the packing scheme determines the number of ciphertexts and plaintexts required to run each ResNet block.
In the SISO-based HCNN filter, the size of filter plaintexts increases by a factor of $w_ih_i$ as each filter element is duplicated to the size of an input image.
The number of slots for weight plaintexts is $w_i h_i f^2 c_i c_o$.
Thus, weight plaintexts significantly outweigh ciphertexts in terms of memory footprint, which only requires $w_i h_i c_i d$ slots.
Figure~\ref{memory study} illustrates the total memory capacity of ciphertexts and plaintexts of each CNN residual block.
In the case of ResNet-20, our implementation shows up to a 14.75\% memory-capacity overhead compared to the baseline.
This overhead is due to the increase in the number of intermediate ciphertexts and bias plaintexts when using CAConv and RAConv, as HyPHEN does not apply PRCR for CIFAR-10.
However, for ResNet-18, our implementation achieves 6.15-6.81$\times$ memory reduction compared to the baseline.

\section{Related Work}
\label{sec:RelatedWorks}
\subsection{HE-based Privacy Preserving Machine Learning}
\label{subsec:HEPPML}
Several variants of homomorphic encryption (HE) have been explored for Privacy-Preserving Machine Learning (PPML).
Early literatures~\cite{aharoni_2020_helayers,lou_2021_hemet,brutzkus_2019_LoLa,gilad_2016_cryptonets,dathathri_2020_eva} employed Leveled HE (LHE), which lacks support managing ciphertext errors and can only perform a limited number of operations, restricting its use to shallow networks. 
Convolutions in these prior works can still be adopted for FHE but are less efficient than convolutions specifically designed for FHE.
For example, \cite{aharoni_2020_helayers} devised a general tensor framework based on tiling, which can serve as an alternative to a SISO-based framework for specific HE parameter settings and image sizes.
We observe that tiling incurs excessive bootstrapping for FHE-based CNNs in general, so we mainly focus on prior FHE CNN implementations for comparison.
By contrast, SHE~\cite{lou_2019_she} implements CNN based on a TFHE scheme~\cite{chillotti_2020_tfhe} and evaluates non-linear functions through table lookups, but it requires a long latency per operation. 
Using RNS-CKKS, \cite{lee_2022_low, lee_2022_access} have demonstrated ResNet implementation on CIFAR-10, employing high-degree polynomial approximation of ReLU.
\cite{lee_2022_low} reported a single thread implementation of ResNet-20, which took 2271 seconds by efficiently utilizing ciphertext slots.
A concurrent work~\cite{kim_2023_optcnn} utilizes coefficient-packed ciphertexts that deploy values as the coefficient of a ring polynomial.
This method reduces the number of rotations in convolution at the cost of limiting flexibility, as a single convolution must be paired with a single bootstrapping.

\subsection{Hybrid PPML}
\label{subsec:Hybrid PPML}
To address the high computational complexity in PPML, the HE-MPC hybrid PI protocol has gained attention as an alternative solution.
In this protocol, client-aided MPC handles non-linear functions such as ReLU while HE operations compute linear functions.  
HE-MPC protocols~\cite{juvekar_2018_gazelle,lehmkuhl_2020_delphi, rathee_2020_cryptflow2,huang_2022_cheetah} have made significant progress. Cheetah~\cite{huang_2022_cheetah} introduced an efficient packing scheme that removed rotations.
However, it is challenging to make a fair comparison between HCNN and HE-MPC protocols due to difference in their security models. HE-MPC protocols also reveal network architecture.
The hybrid approach places less computation burden on the server but assumes 1) continuous network communication and 2) comparable client-side computation power for optimal performance. 
In contrast, FHE only requires succinct communication for transmitting and receiving small input and output ciphertexts. Contrary to the common belief that HE-MPC protocols are significantly faster than FHE-based implementations, our results show comparable performance to prior HE-MPC protocols.

\subsection{FHE hardware acceleration}
\label{subsec:HWacc}
FHE-based applications are promising especially when combined with hardware acceleration. In response to the IT industry's need for privacy-preserving services with realistic quality of service (QoS), prior studies~\cite{jung_2021_access, kim_2020_ntt} have introduced and analyzed the characteristics of FHE operations from a computer architectural perspective. 
This sufficient analysis has led to well-suited solutions for various hardware platforms such as CPU~\cite{boemer_2021_hexl}, GPU~\cite{jung_2021_100x,fan_2023_tensorfhe}, FPGA~\cite{2020_riazi_heax,agrawal_2023_fab}, and ASIC~\cite{kim_2022_bts, kim_2022_ark, nikola_2022_craterlake}. While our HCNN implementation demonstrates performance on CPU and GPU platforms, specialized FHE accelerators can achieve 2--3 orders of magnitude performance improvements.

\section{Discussion}
\label{sec:discussion}
In our implementation, HyPHEN, we introduced several significant enhancements to boost the performance of HCNN.
First, we incorporated a low-degree polynomial activation function obtained from AESPA. Additionally, we developed HCNN based on the GPU implementation of RNS-CKKS. 
Most importantly, we proposed novel algorithms to tackle two key challenges of HCNN: computational complexity and memory footprint.

\subsection{Computational complexity}
We conducted a comprehensive bottleneck analysis of previous HCNN implementations and discovered that a performance limitation arises from FHE rotations. We found out that the majority of the rotation is attributed to the summation of channels within a ciphertext (RaS) and the adjustment of data format between two convolution layers (IR).
To address this challenge, we developed novel algorithms, RAConv and CAConv, which enable encrypted convolution without the need for IR and effectively reduce RaS rotations.
We also proposed a hybrid packing method capable of efficiently managing gaps introduced by strided convolution while minimizing the required rotations.  
Our implementation, HyPHEN, demonstrated a substantial reduction in execution time, from tens of minutes to just a few seconds.

\subsection{Memory footprint}
In addition to the memory expansion resulting from encryption and encoding procedures in the FHE scheme, we identified another significant source of memory expansion: duplicated data in packing methods.
Specifically, while plaintext has lower memory expansion compared to ciphertext, it suffers from additional memory expansion because each weight element occupies the same number of slots as the input image size.
Our experiments on the ImageNet dataset highlighted that loading weight plaintexts from host memory can significantly hinder the performance in CNN models such as ResNet-18. The available High Bandwidth Memory (HBM) in accelerators may not be capable of accommodating the entire model weights.
Our proposed solution, PRCR, addressed this problem by introducing a novel data arrangement that eliminates the need to duplicate each weight element to match the size of an input image. 
For larger models like ResNet-50, the memory reduction ratio achieved by PRCR becomes even more significant. The problem of memory footprint becomes particularly crucial when utilizing specialized accelerators, which are reported to be one or two orders of magnitude faster than GPUs but still share the same HBM technology and, therefore, have the same capacity limitations.

\subsection{Limitation}
\reviewerthree{
While HyPHEN has made significant strides in advancing HCNNs, several limitations should be considered.
Despite achieving a remarkable reduction in execution times, a few seconds for inference are still challenging for deployment in real-world scenarios.
Further, as image classification models evolve to achieve higher accuracy, they require larger memory capacity.
Advanced memory footprint reduction algorithms could further facilitate the broad adoption of PPML.
}
\section{Conclusion}
\label{sec:Conclusion}

In this paper, we proposed HyPHEN, an efficient private inference construction of FHE-based CNN (HCNN).
Combining two convolution methods with reordering and 2D gap packing enables fast inference by significantly reducing the number of homomorphic rotations in convolution.
Additionally, PRCR enables HyPHEN to reduce the memory footprint for high-resolution image classification tasks, which is especially beneficial for memory-constrained devices.
Our experiments with HyPHEN on CPU systems show 1.83$\times$/2.15$\times$ lower latency compared to the prior state-of-the-art algorithm baseline in ResNet-20/18.
Using GPU acceleration, HyPHEN achieves 1.40s/2.17s/2.96s execution time for running ResNet-20/32/44 for CIFAR-10, and we also demonstrated HCNN inference of ResNet-18 for ImageNet in 14.69s for the first time.
We have showcased the practicality of utilizing FHE as a solution to achieve private inference through HyPHEN, which exhibits reasonable execution time while enabling client to benefit from succinct computation and communication processes.

\bibliographystyle{unsrt}
\bibliography{reference}

\appendix
\section{Notations}
\label{sec:Notations}

Table~\ref{tab:notation} tabulates the symbols and their descriptions used throughout the paper.

\begin{table}[htp]
  \caption{Notations and descriptions of the symbols used.}
  \label{tab:notation}
  \centering
  \small
  \begin{tabularx}{\columnwidth}{lX}
    \toprule
    \textbf{Notation}                       & \textbf{Description} \\
    \midrule
    $L$                                     & Maximum level  \\    
    $L^{\prime}$                            & Multiplicative level  \\
    $\ell$                                  & Current level \\
    $n_i, n_o$                              & \# of in and out ciphertexts through convolution\\
    $g$                                     & Width (height) of gap\\
    $s$                                     & Stride of convolution\\
    $pad$                                   & Zero padding of convolution\\
    $m, d$                                  & Gap packing configuration -- a pair of multiplexed channels and duplicates\\
    Dnum                                    &  Decomposition number~\cite{hen_2020_betterbt}\\
    Evk                                     &  Evaluation key, utilized at mul or rotation. \\
    \bottomrule
\end{tabularx}
\end{table}

\end{document}